\documentclass{aa}  

\usepackage{graphicx}
\usepackage{txfonts}
\begin{document}

   \title{Evolutionary phases of merging clusters as seen by LOFAR}
   \titlerunning{Abell 1314, Abell 1319, and Z7215 LOFAR}

   \author{A. Wilber
          \inst{1}\thanks{amanda.wilber@hs.uni-hamburg.de}
          \and
          M. Br{\"u}ggen\inst{1}
          \and
          A. Bonafede\inst{1,2,3}
          \and
          D. Rafferty\inst{1}
          \and
          T. W. Shimwell\inst{4,5}
          \and
          R. J. van Weeren\inst{5}
          \and
          H. Akamatsu\inst{6}
          \and
          A. Botteon\inst{2,3}
          \and
          F. Savini\inst{1}
          \and
          H. Intema\inst{5}
          \and 
          L. Heino\inst{1}
          \and
          V. Cuciti\inst{2,3}
          \and
          R. Cassano\inst{2}
          \and 
          G. Brunetti\inst{2}
          \and
          H. J. A. R{\"o}ttgering\inst{5}
          \and 
          F. de Gasperin\inst{1} 
          \fnmsep
          }
          
        \authorrunning{Wilber et al.}  

   \institute{Hamburger Sternwarte, Universit{\"a}t Hamburg, Gojenbergsweg  112, 21029 Hamburg, Germany
         \and
         Dipartimento di Fisica e Astronomia, Universit\`{a} di Bologna, via P.~Gobetti 93/2, I-40129 Bologna, Italy
         \and
             INAF/Istituto di Radioastronomia, Via P Gobetti 101, 40129 Bologna, Italy
          \and 
          ASTRON, the Netherlands Institute for Radio Astronomy, Postbus 2, 7990 AA, Dwingeloo, The Netherlands
          \and 
          Leiden Observatory, Leiden University, PO Box 9513, 2300 RA Leiden, The Netherlands
          \and 
          SRON Netherlands Institute for Space Research, Utrecht, The Netherlands
             }

   \date{Received --,--; accepted --,--}
 
   \abstract{Massive, merging galaxy clusters often host giant, diffuse radio sources that arise from shocks and turbulence; hence, radio observations can be useful for determining the merger state of a cluster. In preparation for a larger study, we selected three clusters -- Abell 1319, Abell 1314, and RXC J1501.3+4220 (Z7215) -- making use of the new LOFAR Two-Metre Sky Survey (LoTSS) at 120-168 MHz, and together with archival data, show that these clusters appear to be in pre-merging, merging, and post-merging states, respectively. We argue that Abell 1319 is likely in its pre-merging phase, where three separate cluster components are about to merge. There are no radio halos nor radio relics detected in this system. Abell 1314 is a highly-disturbed, low-mass cluster which is likely in the process of merging. This low-mass system does not show a radio halo, however, we argue that the merger activates mechanisms that cause electron re-acceleration in the large 800 kpc radio tail associated with IC~711. In the cluster Z7215 we discover diffuse radio emission at the cluster center, and we classify this emission as a radio halo, although it is dimmer and smaller than expected by the radio halo power versus cluster mass correlation. We suggest that the disturbed cluster Z7215 is in its post-merging phase. Systematic studies of this kind over a larger sample of clusters observed with LoTSS will help constrain the time scales involved in turbulent re-acceleration and the subsequent energy losses of the underlying electrons.}


  \maketitle
%

\section{Introduction}

In the process of the hierarchical formation of structure in the Universe, clusters of galaxies merge to form more massive clusters \citep[see][for a review]{2012ARA&A..50..353K}. Cluster mergers are violent phenomena, releasing large amounts of energy ($\sim 10^{63}$ erg) into the intracluster medium (ICM). A merging system of galaxy clusters will go through three main evolutionary phases in a period of about one gigayear: a pre-merging phase where the galaxy clusters have begun their infall and the intracluster media begin to interact, a merging phase where core passage occurs, and a post-merging phase where the ICM relaxes as shocks and turbulence begin to dissipate. The avenues of this energy release are not well understood and concern fundamental properties of dilute, magnetized plasmas. \\

Customarily, X-ray observations are used to classify clusters of galaxies as either ``merging'' or ``non-merging''. Merging clusters are identified by morphologically disturbed thermal Bremsstrahlung emission from the ICM while non-merging clusters typically have a relaxed morphology and compact cool-cores \citep[e.g.][]{2001ApJ...560..194M}. Several disturbed, massive clusters have been observed to host cluster-scale sources of radio synchrotron emission. Such sources come in two forms: radio halos, which are  diffuse, unpolarised sources that fill most of the inner volume of clusters, and radio relics, which are elongated, polarised sources found in the cluster outskirts. Radio halos and relics typically exhibit a steep synchrotron spectrum ($\alpha < -1$ where $S \propto \nu^{\alpha}$), have low surface-brightness, and are usually on the order of 500 kpc to 1 Mpc in size \citep[see][for a review]{feretti2012}. The origins of these sources and the acceleration mechanisms of the cosmic-ray electrons that power them are still under debate \citep[see][for a review]{2014IJMPD..2330007B}.\\

Radio halos are thought to be powered by turbulent re-acceleration of electrons. In this scenario, turbulence generated by a cluster merger re-accelerates mildly-relativistic electrons \textit{in situ} \citep{2001ApJ...557..560P,2001MNRAS.320..365B,2008Natur.455..944B}. These cosmic-ray electrons then produce synchrotron radiation at radio frequencies within the cluster magnetic field, which is on the order of a few $\mu$G \citep{2002ARA&A..40..319C}. 
The specific radio luminosities of halos observed at 1.4~GHz range between about $10^{23}$ and $10^{25}$~W~Hz$^{-1}$. Halos usually occur in isolation,  
with the exception of the double radio halos in the pre-merging cluster pair Abell~399--401 \citep{2010A&A...509A..86M} and Abell~1758N--A1758S \citep{2018MNRAS.478..885B}. Radio halos  have a smooth  brightness distribution that roughly follows the distribution of the thermal ICM; however, some halos with irregular morphologies have been found \citep[e.g.][]{2009ApJ...704L..54G,2011A&A...530L...5G}.
Some radio halos exhibit exceptionally steep spectra, with $\alpha \sim -2$. These ultra-steep spectrum radio halos (USSRH) may originate from low-mass mergers where the energy budget for turbulent re-acceleration is smaller, or when the radio halo starts to fade at a later evolutionary stage. A prominent example of such a USSRH is the one discovered in Abell~521 \citep{2008Natur.455..944B,2009ApJ...699.1288D}.\\

Questions remain concerning the particle acceleration efficiency of this turbulence, the source of mildly-relativistic seed-electrons that must already fill the ICM before they are re-accelerated, and the origin and amplification of the cluster magnetic field. A handful of unusual clusters are offering interesting insights into these questions. For example, cluster-scale diffuse radio emission has been seen in some cool-core, relaxed clusters \citep[e.g.][]{2014MNRAS.444L..44B}. Furthermore, not all merging systems show radio halos and only upper limits for the diffuse radio flux are known \citep[e.g.][]{2007ApJ...670L...5B}. The fraction of clusters that host radio halos provides fundamental inputs for theory \citep{2010ApJ...721L..82C}. Radio relics are more direct indicators of merger activity as they are thought to trace merger-induced shocks. The position of the radio relics with respect to other cluster components, such as galaxies and Dark Matter, can help to reconstruct the geometry of the merger \citep[e.g.][]{2018arXiv180610619G}. \\ 

Another way to assess the merging state of clusters is to examine the interplay between the radio lobes of active galactic nuclei (AGN) and the intracluster medium. Active radio galaxies are suspected to contribute a population of relativistic electrons to their surrounding medium, which may provide a portion of the seed-electrons needed to explain diffuse ICM radio sources. An example of such is the connection between a radio relic and a radio galaxy in Abell 3411-3412 \citep{2017NatAs...1E...5V}. Extended radio galaxies, with sizes $\geq~700$ kpc, may serve as large suppliers of these seed-electrons \citep[e.g.][]{2018MNRAS.473.3536W}. Injection of seed-electrons from AGN can also be studied by looking at remnant radio galaxies, which are radio galaxies in the phase after which nuclear activity has ceased \citep[e.g.][]{2017A&A...606A..98B, 2014IJMPD..2330007B}. For radio galaxies with intermittent activity, this remnant, or fossil, emission may appear as fading lobes which are visibly detached from an active AGN core. Radio galaxies moving through the cluster environment may also experience ram pressure that results in a bent-tail (BT) morphology \citep{1980ARA&A..18..165M}. Studying the output of radio AGN within cluster environments is crucial to understanding the characteristics of the ICM. \\

In recent years, the LOw Frequency ARray (LOFAR) has been a key instrument used to study faint cluster-scale radio sources associated with merging galaxy clusters. LOFAR is a low-frequency radio interferometer with a compact core in the Netherlands and stations presently located in six other European countries \citep{vanHaarlem2013}. The LOFAR Two-Metre Sky Survey (LoTSS; \citealp{2017A&A...598A.104S}) is a new low-frequency survey aiming to map the entire northern sky. When completed, it will be two orders of magnitude deeper in point-source sensitivity and one order of magnitude higher in resolution than any current very large radio survey at this frequency regime (Shimwell et al. submitted 2018). The first 600 pointings (20\% of the total) of this survey have recently been observed, covering the 120--168~MHz band using the high-band antennas (HBA). LOFAR has already yielded valuable insights into radio halos. These include the observation of intricate and filamentary structures embedded in a radio halo \citep{2016MNRAS.459..277S}, the discovery of a new USSRH \citep{2018MNRAS.473.3536W}, the detection of extended radio emission in two cool-core clusters \citep{Savini2018c}, and the observation of a tentative bridge of emission connecting two radio halos \citep{2018MNRAS.478..885B}.\\

Theoretical studies have predicted that LOFAR will detect many more radio halos, especially halos with ultra-steep spectra ($\alpha \leq -1.5$) since these objects appear brighter at lower frequencies \citep[e.g.][]{2008Natur.455..944B,2012A&A...548A.100C}. Studies of galaxy clusters made with observations from the Giant Metrewave Radio Telescope (GMRT), operating in a frequency range of 150 MHz to 1.4 GHz, have already revealed that radio halos can change in brightness and morphology depending upon the observing frequency, and that some radio halos are only visible at lower frequencies \citep[e.g.][]{2007A&A...463..937V,2008A&A...484..327V,2015A&A...579A..92K}.  \\

Here we report on our search for cluster-scale diffuse radio emission associated with three merging cluster systems: Abell 1314, Abell 1319, and RXC J1501.3+4220 (referred to as Z7215 hereafter). In preparation for a larger study, which will consist of LOFAR observations of many merging clusters, we selected these three clusters since they were already covered by LoTSS and appeared to host potential diffuse emission upon first inspection of preliminary images. These three clusters represent a small set of lower-mass mergers that are likely at different evolutionary stages. We aim to determine the nature of the radio emission in these clusters and relate this to their merger phase. The following cosmology, H$_{0}=69.6$, $\Omega_{m}=0.286$, and $\Omega_{\Lambda}=0.714$, is used hereafter.


\section{Methods}

\subsection{LOFAR observations and data reduction}
The 8-hr LOFAR observations we report in this paper were made as part of LoTSS \citep{2017A&A...598A.104S} over a frequency range of 120-168 MHz using the Dutch HBAs. The data reduction steps for these data are identical to the steps detailed in \citet{2018MNRAS.473.3536W, 2018MNRAS.476.3415W}, but are also re-iterated below.  

\subsubsection{Prefactor}

Prefactor\footnote{\url{https://github.com/lofar-astron/prefactor}} is a package containing automated pipelines called Pre-Facet-Calibration and Initial-Subtract. Pre-Facet-Calibration compresses and averages the original data and performs the initial direction-independent calibration (see \citet{deGasperin2018a} for details). In this step a flux calibrator (observed at the beginning and end of the target observation) was used to compute amplitude gain solutions, station clock offsets, station phase offsets, and station differential total electron content (dTEC). Amplitude gain solutions and corrections for clock and phase offsets were then transferred to the target field data. An initial phase calibration was also performed using a global sky model from the TIFR GMRT Sky Survey (TGSS) at 150 MHz \citep{2017A&A...598A..78I}. For the observations of all three of our selected clusters, the calibrator 3C196 was used, a bright quasar (66 Jy at 159 MHz according to the \citealt{2012MNRAS.423L..30S} absolute flux scale). After the direction-independent calibration was completed, preliminary imaging was carried out via the Initial-Subtract pipeline. The full wide-field of the calibrated target data was imaged in high and low resolution using {\tt WSClean} \citep{2014MNRAS.444..606O}. These full field images were used to model and subtract all sources in preparation for direction-dependent calibration. 

\subsubsection{FACTOR \label{Facet Calibration}}
Direction-dependent calibration for our LoTSS data was carried out through the facet calibration technique \citep{vanWeeren16}. This method of calibration is executed via the FACTOR\footnote{\url{http://www.astron.nl/citt/facet-doc/}} software package. FACTOR tesselated the full target field into several smaller portions of sky called facets, where each facet is automatically chosen to be centered on a bright compact source to be used as a facet calibrator. TEC, phase, and amplitude solutions were computed from the facet calibrator and applied to all the sources in that facet. Facets were processed in order of brightness, and all facet sources were subtracted from the $uv$-data before processing the next facet. This method reduced the effective noise in subsequent calibration steps. The target facet, containing the cluster of interest, was appointed as the last facet in the processing list, such that all other nearby and bright sources were already calibrated and subtracted. The calibration region for the target facet was chosen either as the cluster center, containing multiple sources, or as a single bright, compact source near the cluster center. For more details on facet calibration the reader is referred to \citet{vanWeeren16}, \citet{shimwell2016}, and \citet{2016MNRAS.460.2385W}.\\

\subsection{LOFAR imaging}
The final FACTOR-calibrated data of all three clusters were imaged and analysed with CASA tools (Common Astronomy Software Applications; \citealp{2007ASPC..376..127M}). To lower the resolution and increase the sensitivity to diffuse radio emission on large scales, we used the CASA task CLEAN and chose an increased outer $uv$-taper of up to 30 arcsec and Briggs' robust values of -0.25 or 0. Final images made in CASA were also corrected for the LOFAR station beam. The error in our flux density measurements are assumed to be 10\%, which was determined by comparing flux densities of several sources in our LOFAR map to the same sources in TGSS \citep{2017A&A...598A..78I} and the 7C survey \citep{2007MNRAS.382.1639H}. In the following subsections we describe the basic steps for data manipulation and imaging that were made on our LOFAR measurement sets.

\subsubsection{Subtraction of compact sources\label{uplimcalc}}

To accurately measure diffuse radio emission at the cluster center, we used an image made after subtracting compact radio sources. Compact sources are usually found within the vicinity of the cluster center, and are typically associated with AGN. Hence, the scale of the compact emission we wish to subtract is on the order of $<200$ kpc. A compact-source image was made using the task CLEAN in CASA by selecting a specific $uv$-range which corresponds to emission on a projected scale of $>200$ kpc. Based on the redshift of the individual clusters, the $uv$-cut was made at $800\lambda$ for Abell 1314, and $4500\lambda$ for Abell 1319 and Z7215. The model components from this compact-source image were then subtracted from the measurement set using tasks FTW and UVSUB. We then re-imaged the source-subtracted dataset with a $uv$-range of $> 80\lambda$, and used an increased outer $uv$-taper of 30 arcsec to bring out diffuse emission. \\

If diffuse emission was present at the cluster center, we measured the flux density of this diffuse emission contained within $2\sigma$ or $3\sigma$ contours. From this flux density we calculated a power at 144 MHz considering the redshift of the cluster, and we extrapolated this power to 1.4 GHz with a spectral index of $\alpha = -1.3$. This power at 1.4 GHz is used to give a relative comparison to the power in the correlation for radio halo power versus cluster mass (referred to hereafter as the $P-M$ correlation) from \citet{2013ApJ...777..141C}.

\subsubsection{Injection of mock halo}

For the massive, merging cluster Z7215, we followed a previously used method of injecting a fake radio halo into the visibilities. This method has been typically used to compute upper limits to the diffuse emission of clusters without radio halos \citep[e.g.][]{2008A&A...484..327V,2013A&A...557A..99K,2015A&A...579A..92K,2017MNRAS.470.3465B}. These studies showed that the flux density of the injected radio halo is only partly recovered and this effect becomes relatively more important for fainter halos. Since we detected faint diffuse emission at the center of Z7215 (see section \ref{Z7215res}) we used the injection procedure to evaluate the effect of such losses in the measurement of this diffuse flux \citep[e.g.][]{2007ApJ...670L...5B, 2008A&A...484..327V,2017MNRAS.470.3465B,2018A&A...609A..61C}. Although Abell 1314 is a highly-disturbed cluster (as seen by X-ray observations), detectable emission is not expected in such a low-mass system (M$_{500} < 10^{14}$ M$_{\odot}$) owing to energy limitations \citep[e.g.][]{2005MNRAS.357.1313C, 2006MNRAS.369.1577C, 2010A&A...517A..10C, 2016MNRAS.456.1259B}. Therefore, we decided not to implement halo injection for Abell 1314. \\

Following these methods, we simulated a radio halo by injecting a radio source into the $uv$ data that has a central brightness $I_{\rm 0}$ and an $e$-folding radius of $r_{e}$ \citep{2009A&A...499..679M}. The length scale, $r_e$, is defined as the radius at which the brightness of a radio halo drops to $I_{\rm 0}/e$, and measuring this is independent on the sensitivity of the radio images\footnote{\citet{2017MNRAS.470.3465B} compared the values of RH and re found by \citet{2007MNRAS.378.1565C} and \citet{2009A&A...499..679M} for the eight clusters in common in their samples and found that the median value of the ratio R$_H$/r$_e$ is 2.6.}. The model of this mock radio halo was Fourier transformed into the visibility data (MODEL\_DATA column), taking into account the w-projection parameter \citep{2005ASPC..347...86C}. A relatively empty region near the cluster centre, void of bright sources and artefacts, was chosen to host the injected flux. The dataset was then re-imaged with an outer $uv$-taper of 30 arcsec. We considered the injected halo detected when it was recovered above $2\sigma$ contours with a diameter of roughly $3r_e$.\\

\subsection{Supplementary observations }

\subsubsection{GMRT observations of Abell 1314} 
Archival GMRT data at 235 and 610 MHz (Obs ID: 7909) centered on the cluster Abell 1314 were obtained and re-processed with the SPAM pipeline \citep[see][for details]{2017A&A...598A..78I}. We used the 610 MHz data with our LOFAR data at 144 MHz to produce a spectral index map of the sources in Abell 1314\footnote{The 235 MHz image was not useful for our multi-frequency analysis.}. To produce the spectral index map, we imaged the cluster at both frequencies using the same clean settings in CASA clean: uvrange $> 200 \lambda$, uniform weighting, and a $uv$-taper of 40 arcsec. We re-grided the LOFAR map to the 610 MHz GMRT map and smoothed the images to give the same beam size (47 arcsec $\times$ 35 arcsec).
 
\subsubsection{X-ray observations}\label{sec:xmm_data}

Shallow \textit{XMM-Newton} observations (Obs ID: 0149900201 and 0402250201, raw exposure times = 18 and 27.2 ksec, respectively) were used to characterize the thermal ICM emission in Abell 1314 and Z7215, respectively. The SAS v15.0 and the built-in extended source analysis software (ESAS) were used to process and calibrate the data obtained with the \textit{XMM-Newton} European Photon Imaging Camera (EPIC). Following a standard procedure, the raw data were created by \textit{emchain}, and the light curves were extracted and screened for time-variable background components by the \textit{mos-filter, pn-filter} task. The total exposure time after screening was 16.9 ks, 17.0 ks and 12.9 ks for MOS1, MOS2 and pn, respectively, for the Abell 1314 (0149900201) observation. The total exposure time after screening was 25.6 ks, 25.4 ks and 21.6 ks for MOS1, MOS2 and pn, respectively, for the Z7215 (0402250201) observation. \\

Abell 1319 was observed with Chandra ACIS-S for 9.91 ks (Obs ID: 11760) and Z7215 was observed with the Chandra ACIS-I VFAINT mode for 13ks (Obs ID: 7899). We processed the Chandra data following \citealp{Vikhlinin2005}\footnote{We used CIAO v4.6 and CALDB v4.7.2.}. The background was examined in the 0.5--7~keV band, to search for high proton flares. For the final exposure corrected image we used a binning of 1 arcsec pixel$^{-1}$. For more details the reader is referred to \citet{Vikhlinin2005}. \\

\begin{figure}
\centering
\includegraphics[width=0.49\textwidth]{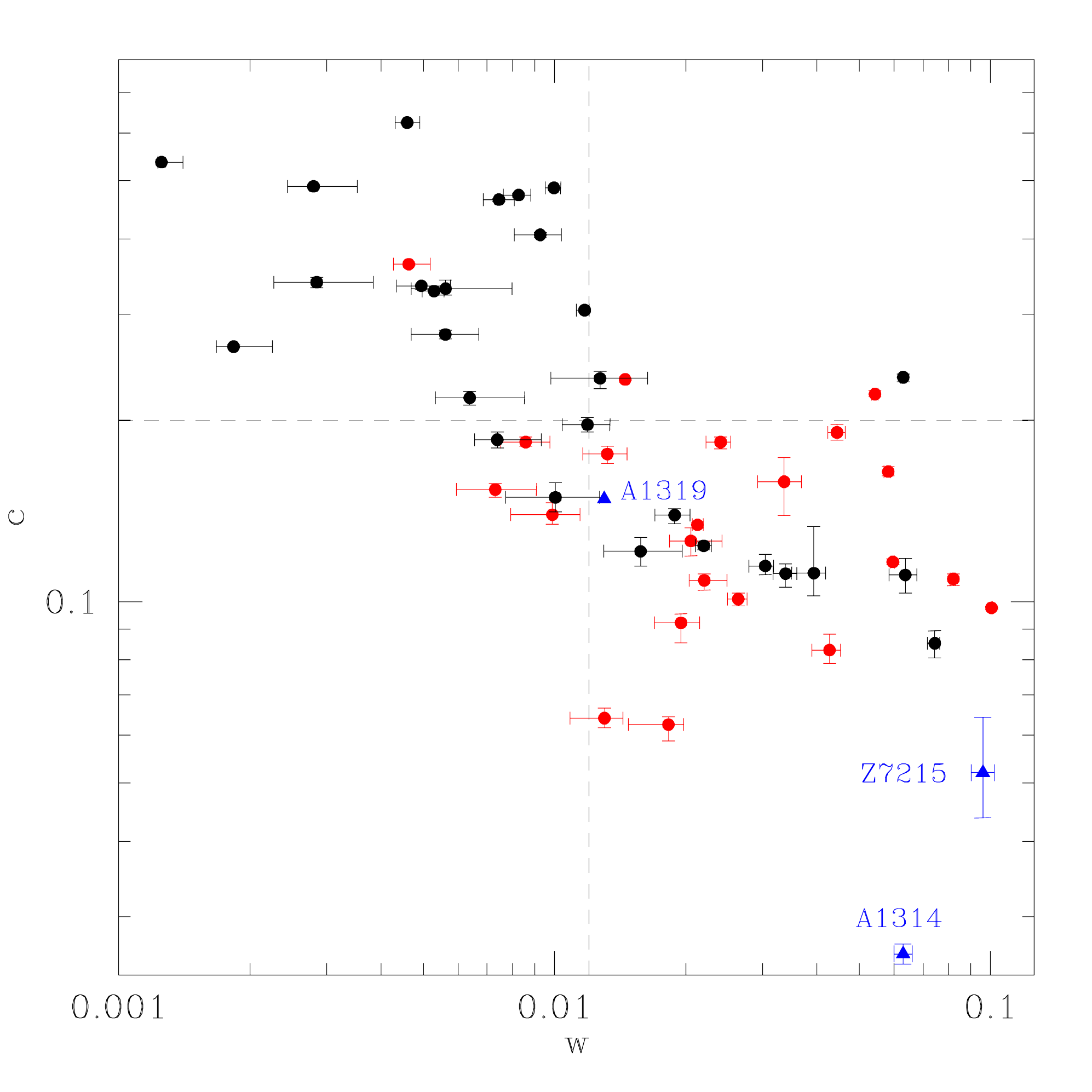}
\caption{A sample of clusters with masses $>6 \times 10^{14}$ M$_{\odot}$ plotted by their concentration parameter $c$ versus their centroid shift $w$ from \citet{2015A&A...580A..97C}. Red dots are clusters hosting radio halos and black dots are clusters without radio halos. We include three blue points for Abell 1314, Abell 1319, and Z7215. The $w$ and $c$ parameters of all clusters are computed from Chandra X-ray images, except for Abell 1314 and Z7215, which have parameters computed from XMM-Newton images. Our selected clusters have masses lower than the sample from \citet{2015A&A...580A..97C}. \label{wc-diagram}}
\end{figure}

By analyzing the surface brightness of the X-ray thermal emission of the ICM, we computed morphological parameters indicative of the merging status of the cluster. The centroid shift parameter, $w$, is defined as the projected separation between the peak and centroid of the X-ray surface brightness distribution when the aperture used to compute it decreases from the aperture radius, $R_{ap} = 500$~kpc, to smaller radii \citep{Bo2010}. The concentration parameter, $c$, is defined as the ratio of the X-ray surface brightness within a radius of 100 kpc over X-ray surface brightness within a radius of 500 kpc \citep{Santos2008}. The c-w morphological diagram adapted from \citet{2015A&A...580A..97C} is shown in Fig.~\ref{wc-diagram}. $c$ and $w$ are anticorrelated, with relaxed clusters lying on the top left region and merging clusters lying on the bottom right region. The boundary lines are taken from \citet{2010ApJ...721L..82C} and defined at $w \sim 0.012$ and $c \sim 0.2$, such that larger values of $w$ and smaller values of $c$ indicate a disturbed, merging cluster \citep{2016A&A...593A..81C}. Our computed values for $w$ and $c$ for each cluster are listed in Table~\ref{tab1} and shown in the diagram in Fig.~\ref{wc-diagram}

\begin{table*}
 \centering
   \caption{Our selected clusters observed by LOFAR. The mass listed for Abell 1319 and Z7215 are SZ mass-estimates from Planck \citep{2014A&A...571A..29P}. The last three columns give the characteristics of a radio halo falling on the $P-M$ correlation from \citet{2013ApJ...777..141C} for a cluster corresponding to that mass. We use these correlation values to compare to the diffuse emission detected by LOFAR. (1): Our mass-estimate from \textit{XMM-Newton} observations and scaling relations from \citet{2011A&A...535A...4R}. Since this cluster is at such a low-mass, the $P-M$ correlation is not applicable. (2): $w$ and $c$ parameters were computed for Abell 1319-A only.}
   \label{tab1}
  \begin{tabular}{ccccccccc}
  \hline
  \hline
Cluster Name & RA, dec & Redshift & Scale & Mass & Dynamics & Halo P$_{1.4~\rm corr}$ & R$_{\rm H~corr}$ & r$_e$  \\
(Obs ID) & J2000 & $z$ & kpc arcsec$^{-1}$ & [$10^{14}$ M$_\odot$] & $w_{500~\rm kpc}$, $c_{100~\rm kpc}$ & log$_{10}$~[W Hz$^{-1}$] & [kpc] & [kpc] \\
\hline
& & & &  &  &  &  &  \\
Abell 1314 & 11h34m50.5s, & 0.0335 & 0.672 & 0.68$^{(1)}$ & $0.063 \pm 0.003$, & N/A & N/A & N/A \\ 

(L229509) & +49d03m28s& & & Low-Mass &  $0.026 \pm 0.01$ &  & &  \\ 
& & & &  &  &  &  &  \\
Abell 1319 & 11h34m13.2s, & 0.2906 & 4.395 & 4.79$^{+0.51}_{-0.49}$ & 0.013, & 23.80 & 351 & 135  \\ 
(L403936)& +40d02m36s & & & Massive & 0.148$^{(2)}$ &  &  &  \\
& & & &  &  &  &  &  \\
RXC J1501.3+4220 & 15h01m23.0s, & 0.2917 & 4.406 & 5.87$^{+0.42}_{-0.41}$ & $0.096 \pm 0.06$, & 24.13 & 422 & 162  \\
-- Z7215 -- & +42d20m40s & & & Massive & $0.052 \pm 0.09$ &  &  &  \\ 
(L371804)& & & &  &  &  &  &  \\
\hline

\end{tabular}
\end{table*}

\begin{figure*}
\centering
\includegraphics[width=\textwidth]{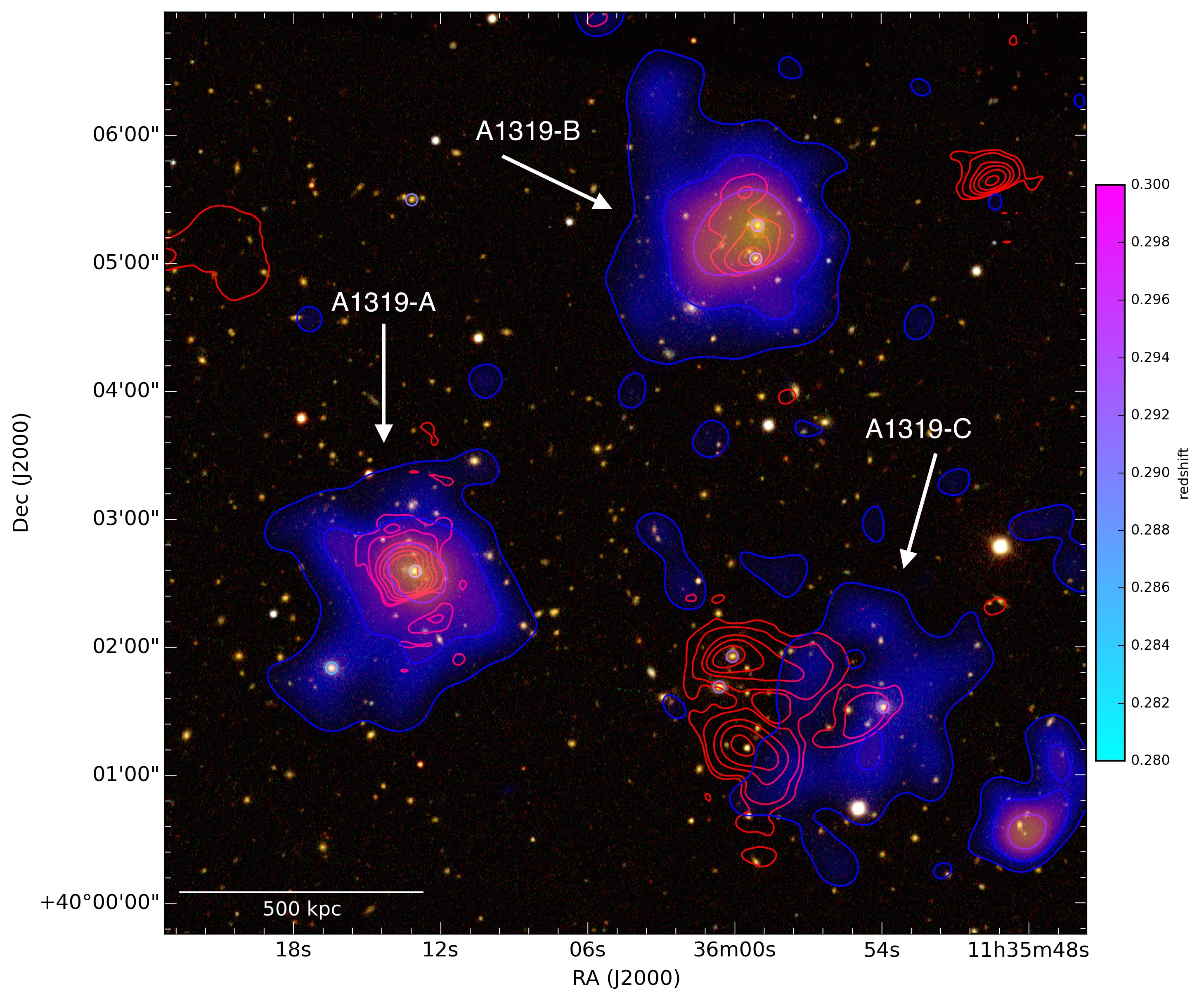}
\caption{SDSS {\it i,r,g} image of Abell 1319 with LOFAR radio emission overlaid as contours. LOFAR high-resolution emission (11 arcsec $\times$ 6 arcsec) is shown by red contours $[3,5,16,64,128,256,512]~\times~\sigma$ where $\sigma = 170~\mu$Jy beam$^{-1}$. The smoothed Chandra X-ray image is shown in blue-magenta and as blue contours. Cluster galaxies with known spectroscopic redshifts are marked by colored rings, where the color represents redshift.  \label{A1319opt}}
\end{figure*}

\begin{figure}
\centering
\includegraphics[width=0.49\textwidth]{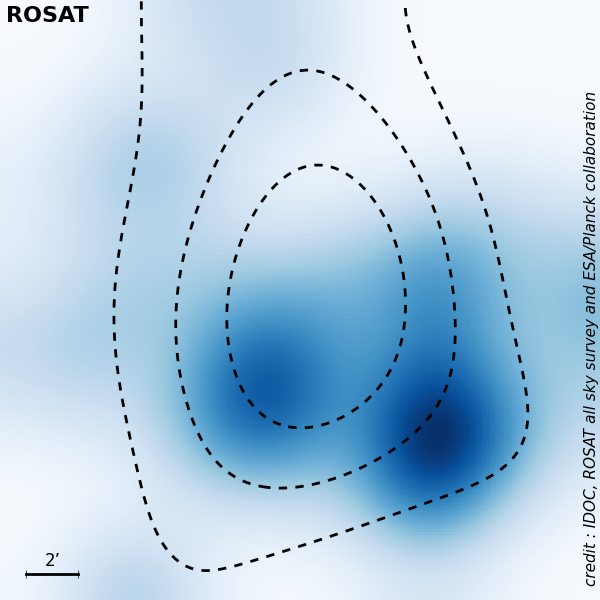}
\caption{This image is taken from the Planck SZ Cluster Database. Abell 1319-A is accompanied by two additional cluster components of similar size at a similar redshift. We suggest the SZ-derived mass is a combination of three low-mass cluster components (faintly visible in blue, as seen by ROSAT).\label{SZ-ROSAT}}
\end{figure}

\begin{figure*}
\centering
\includegraphics[width=\textwidth]{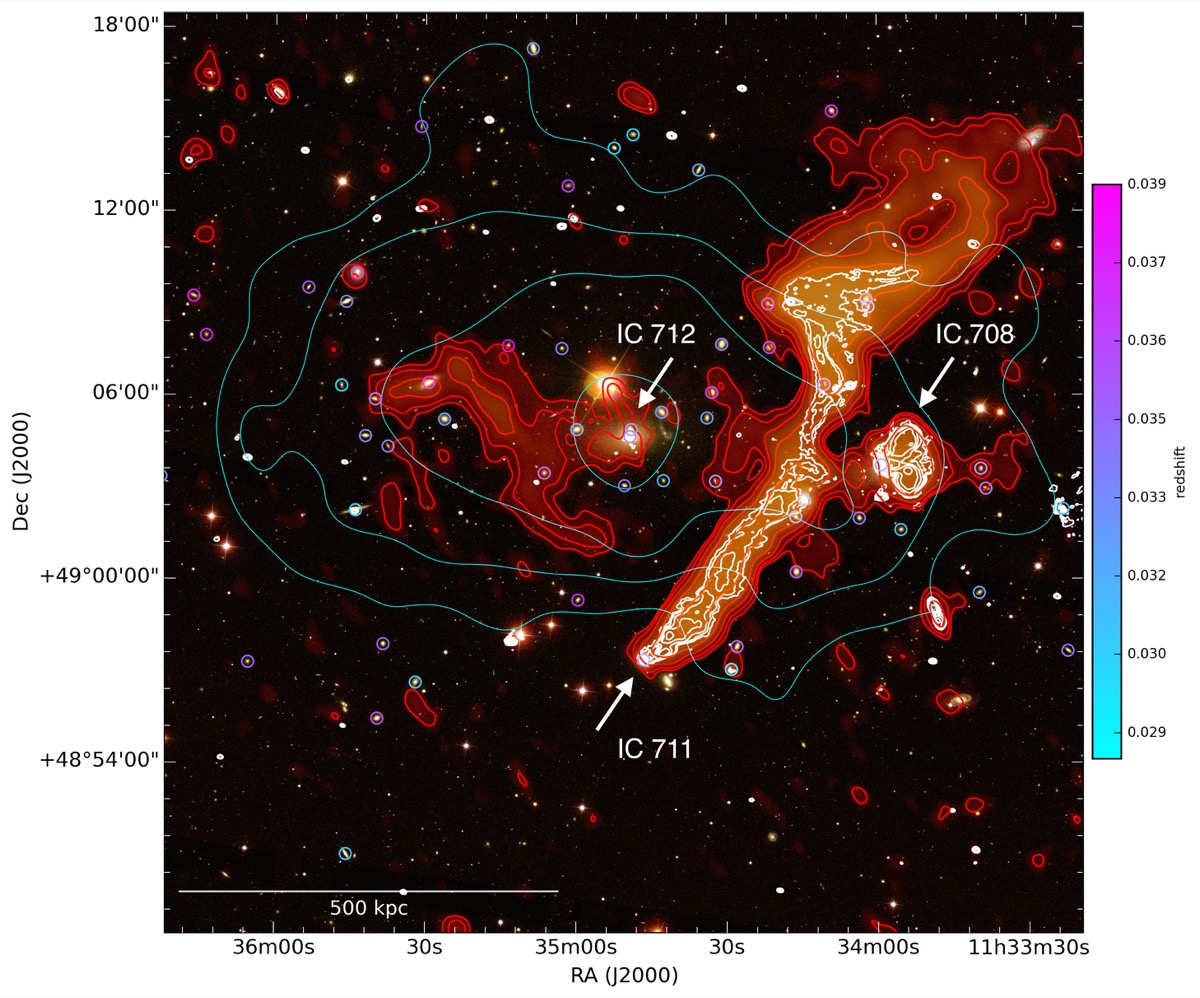}
\caption{SDSS {\it i,r,g} image of Abell 1314 with radio emission overlaid as contours. LOFAR high-resolution (8 arcsec $\times$ 5 arcsec) emission is shown by white contours $[6, 12, 24, 48, 96, 192, 384, 768]~\times~\sigma$ where $\sigma = 180~\mu$Jy beam$^{-1}$. LOFAR diffuse emission after compact source subtraction (with a resolution of 43 arcsec $\times$ 33 arcsec) is shown in red color and red contours where contours are $[3, 6, 12, 24, 48]~\times~\sigma$ and $\sigma=300~\mu$Jy beam$^{-1}$. Smoothed XMM Newton X-ray contours as also shown in cyan. Cluster galaxies with known spectroscopic redshifts are marked by colored rings, where the color represents redshift. \label{A1314opt}}
\end{figure*}

\begin{figure}
\centering
\includegraphics[width=0.49\textwidth]{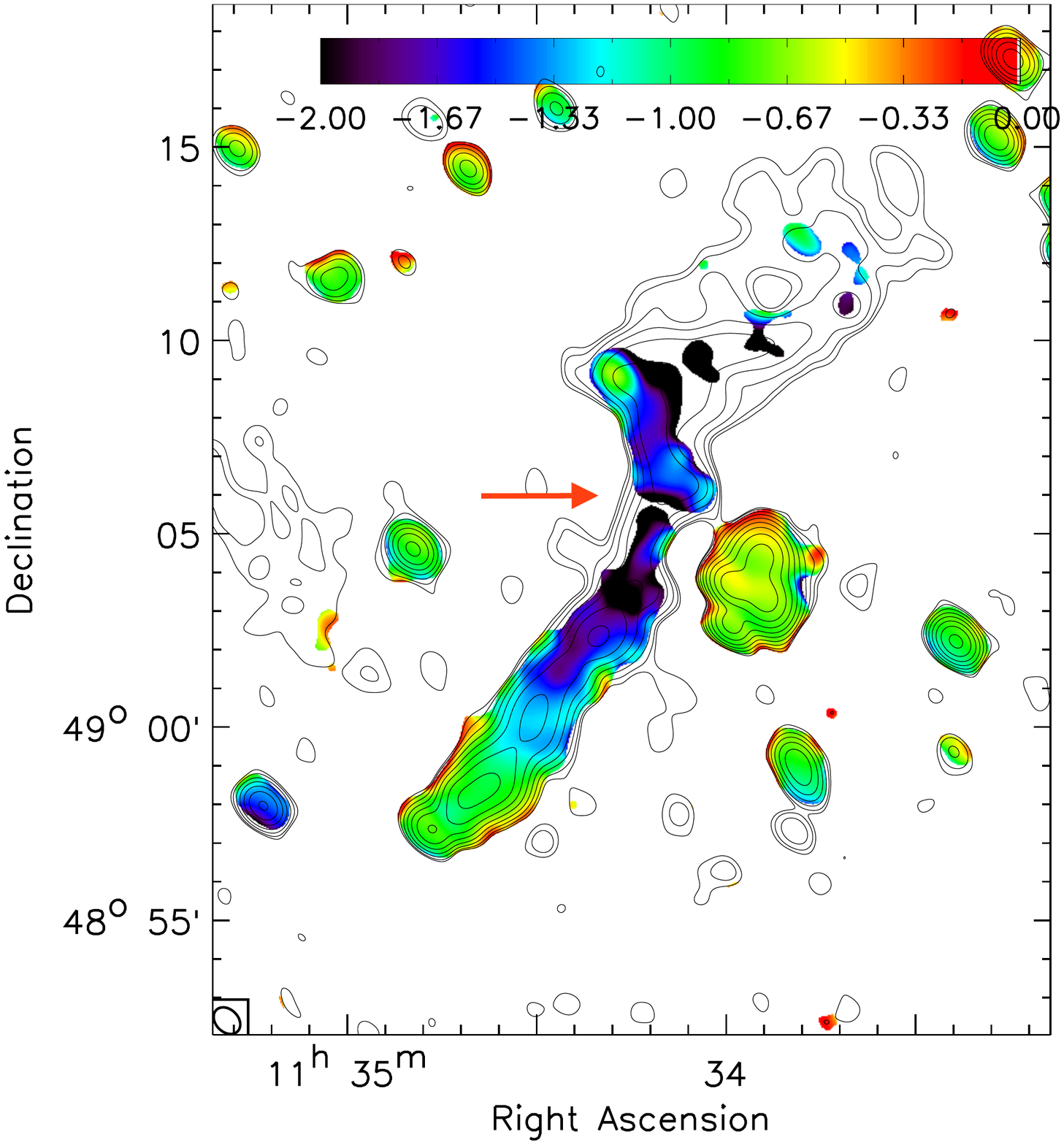}
\caption{Spectral index map over the bent-tail galaxies in Abell 1314. The map is made with the LOFAR map at 144 MHz and the GMRT map at 610 MHz. Both maps are imaged with the same setting and convolved to the same beam size. We indicated no RA, dec offset between the two maps. The radio head-tail from IC~711 shows steepening in the length of the tail with a portion of flattening in the mid-section. A red arrow roughly marks where the flattening of the spectral index occurs. \label{spix}}
\end{figure}

\section{Results}

Abell 1319 and Z7215 were detected by the Planck satellite \citep{2014A&A...571A..29P} via the Sunyaev-Zel'dovich (SZ) effect and have SZ-derived mass estimates (listed in Table~\ref{tab1}). Abell 1314 was not detected by Planck, but the most recent mass-estimate of Abell 1314, from MCXC (Meta-Catalogue X-ray galaxy clusters; \citealt{2011A&A...534A.109P}), is $M_{500} = 0.4608 \times 10^{14}$ M$_{\odot}$, suggesting that this cluster has a very low-mass. We report new radio emission detected by LOFAR in the low-mass cluster Abell 1314, and we derive a mass estimate from $XMM-Newton$ X-ray observations. We also report new diffuse radio emission detected by LOFAR in the intermediate mass cluster Z7215. Table~\ref{tab1} lists the individual details for our three selected clusters. In this paper we hypothesize the merging status of each cluster based on a joint radio-X-ray analysis and suggest that each cluster represents one of the three evolutionary phases of merging systems: Abell 1319 in a pre-merging phase, Abell 1314 in a merging phase, and Z7215 in a post-merging phase.

\subsection{Abell 1319: Example of a pre-merging system}

In Fig.~\ref{A1319opt} we present an overlay of Chandra X-ray and LOFAR radio emission on a Sloan Digitial Sky Survey (SDSS) optical image of the cluster Abell 1319. This image of the cluster shows three ICM components. The main component is Abell 1319-A, and the two additional ICM components, to the northwest (Abell 1319-B) and southwest (Abell 1319-C), show coincident galaxies with spectroscopic redshifts (from the SDSS database) similar to that of the galaxies in Abell 1319-A. \\ 

In Fig.~\ref{SZ-ROSAT} Planck contours are overlaid on a X-ray image of Abell 1319 from ROSAT. All three ICM clumps are included in the Planck contours. Therefore, it is likely that the SZ-derived mass-estimate from Planck (M$_{500} = 4.79^{+0.51}_{-0.49} \times 10^{14}$ M$_{\odot}$) is the mass of the total system, including all three ICM components. Although the Chandra image shows the southwest clump, Abell 1319-C, as being the most faint, ROSAT shows this component to be the brightest of the three, but this effect could be due to differences in the pointing and exposure times of the two separate observations. Abell 1319-C also lies near the CCD edge of the Chandra observation. We did not estimate the mass values of each individual cluster component. \\

Our LOFAR image of this multi-component system shows compact radio emission at the center of each ICM. These are likely to be the radio AGN of bright central galaxies (BCG). The optical overlay shows that the galaxies coincident with these compact central radio sources are all at similar redshifts. A more extended radio galaxy is seen on the eastern edge of Abell 1319-C. This emission is likely a bent-tail radio galaxy from within the cluster (north) and a background radio galaxy (south) appearing in projection to be one extended source. After compact-source subtraction and uv-tapering to lower the resolution of our LOFAR image, no diffuse sources appear to be present at the ICM centers, nor on the ICM edges. We do not estimate upper limits for diffuse emission at the center of each ICM component because residual AGN emission from the BCGs is too prominent for a reliable subtraction. We also cannot compare the residual emission to the $P-M$ correlation since we do not have the individual mass estimates for each clump, and each clump likely has a mass lower than cluster masses tested in the $P-M$ correlation.  \\

Since the active BCGs all appear to reside in their respective ICM centers, and since no radio halos or radio relics are detected in this system, we suggest that this multi-component system is in a pre-merging state, i.e. the separate ICM components are in gravitational in-fall, but have not yet merged. The centroid-shift and the concentration parameter for Abell 1319-A, the eastern clump, also indicate that this ICM component is more relaxed, according to the boundaries on the $w-c$ diagram (with $w = 0.013$, and $c = 0.148$ from Fig.~\ref{wc-diagram}; also see \citealt{2010ApJ...721L..82C,2015A&A...580A..97C, 2016A&A...593A..81C}). These parameters are measured from our Chandra X-ray image of Abell 1319.

\subsection{Abell 1314: Example of a merging system}

Abell 1314 has been studied for several decades with radio telescopes, starting with \citet{1974PASP...86..223W}. There are four IC galaxies in the cluster with radio counterparts. Two bent-tails, associated with IC~708 and IC~711, were identified in 1976 with the Westerbork telescope at 610 MHz and 5 GHz \citep{1976Natur.259..451V}. IC~708 exhibits a wide-angle opening between two tails, both with short extents and radio lobes to the west of the host. IC~711 is a head-tail, or narrow-angle tail, with a single, long radio tail extending hundreds of kpc north of the host galaxy.  \\

\citet{2016arXiv161007783S} presented GMRT observations at 235, 610, and 1300 MHz of the head-tail IC~711 in Abell 1314. They produced a spectral index map over the body of the emission and report a break in the spectrum. They suggest that the morphology of the tail did not form solely from N-S movement from the host galaxy, because there is a sharp turn in the emission at the northernmost extent. More recently, \citet{2017AJ....154..169S} presented a low-resolution image from a 240 MHz GMRT observation of the radio tail associated with IC~711. In this image, it is apparent that there is more emission extending westward at the northernmost region referred to by \citet{2016arXiv161007783S}. \\

An overlay of our LOFAR images of Abell 1314, in high- and low-resolution, can be seen in Fig.~\ref{A1314opt}. The LOFAR high-resolution (8 arcsec $\times$ 5 arcsec) image of Abell 1314 reveals extended tail emission from IC~711. The wide-angle tail IC~708 is also visible to the west of the elongated head-tail. The sharp turn noted by \citet{2016arXiv161007783S} actually appears to be filamentary emission sweeping westward for $\sim 300$ kpc. As noted by \citet{2016arXiv161007783S}, this northernmost emission that appears, in projection, perpendicular to the length of the tail likely did not form from the trajectory of the active host galaxy. It is more likely that the radio emission in the northern part of the tail has been disturbed by the ICM, such as ram pressure from turbulence or shocks traveling from the cluster center outward and toward the West. The LOFAR low-resolution map (shown as red contours and color in Fig.~\ref{A1314opt}) reveals an additional 300 kpc of bulk diffuse emission beyond this high-resolution filamentary portion of the tail, giving a total projected size of 800 kpc. \\

In our low-resolution LOFAR image, faint diffuse emission is also detected in the central regions of the cluster, which has not been detected at higher frequencies. This emission is not spherically uniform in shape or brightness, but instead exhibits ``arms''. It is possible that this emission is remnant AGN emission from the central BCG (IC~712: $z=0.033553$, which shows a compact core), or from other radio galaxies within the cluster center. \\

\subsubsection{Spectral index on radio emission \label{secspec}}

A spectral index map is generated along the head-tail radio galaxy IC~711 by comparing the GMRT map at 610 MHz to our LOFAR map at 144 MHz. The spectral index along the tail steepens from south to north, but there is a region of flattening that takes place in the central part of the tail. The index goes from $\alpha \sim -2$ to $ -1.3$ and then back to $\sim -2$ within this region. A compact source is also seen to the northeast edge of the tail emission, marked by a flat spectral index, but this is likely to be a foreground source. The spectral index in the northernmost portion of the tail, where emission sweeps westward, cannot be determined since this emission is not detected by the archival GMRT observations we obtained at 610 nor 235 MHz. \\

To determine an upper limit on the spectral index of the diffuse emission at the cluster center, we compare the flux density within $3\sigma$ contours in our 144 MHz LOFAR map to the same region of the 610 MHz GMRT map\footnote{We compare the images made with the same clean settings including uniform weighting and the same beam size.}. This flux density within this region is only at noise level at 610 MHz, so we integrate the noise ($300~\mu$Jy beam$^{-1}$) over the area of the $3\sigma$ contour region from the LOFAR map and find the upper limit on the flux density to be 19.4~mJy at 610 MHz. Comparing the measurements gives a spectral index upper limit estimate of $\alpha < -1.3 $.

\subsubsection{Central diffuse radio emission}

The non-spherical, limb-like emission at the center of Abell 1314, with a largest linear size of $\sim~380$ kpc, is probably not related to a cluster radio halo. The flux density of the central diffuse emission is measured in our compact-source-subtracted image made with a $uv$-taper of 30 arcsec. The flux density within $3\sigma$ contours where $\sigma = 400~\mu$Jy beam$^{-1}$ is $109 \pm 11 $ mJy. This translates to a power of $(2.85 \pm 0.29) \times 10^{23}~$W Hz$^{-1}$ at 144 MHz and $(1.48 \pm 0.15) \times 10^{22}~$W Hz$^{-1}$ at 1.4 GHz if extrapolated assuming a spectral index of $\alpha = -1.3$. This power of $\log_{10}$(P$_{1.4}$ $ /$ W Hz$^{-1}) =22.17$ is much lower than the power of radio halos even at the lowest end of the cluster mass range (see the $P-M$ correlation from \citet{2013ApJ...777..141C} and \citet{2016A&A...595A.116M}). As we discuss in Sec.~\ref{disA1314}, it is more plausible that this emission is a remnant radio galaxy.   \\

\begin{figure}
\centering
\includegraphics[width=0.49\textwidth]{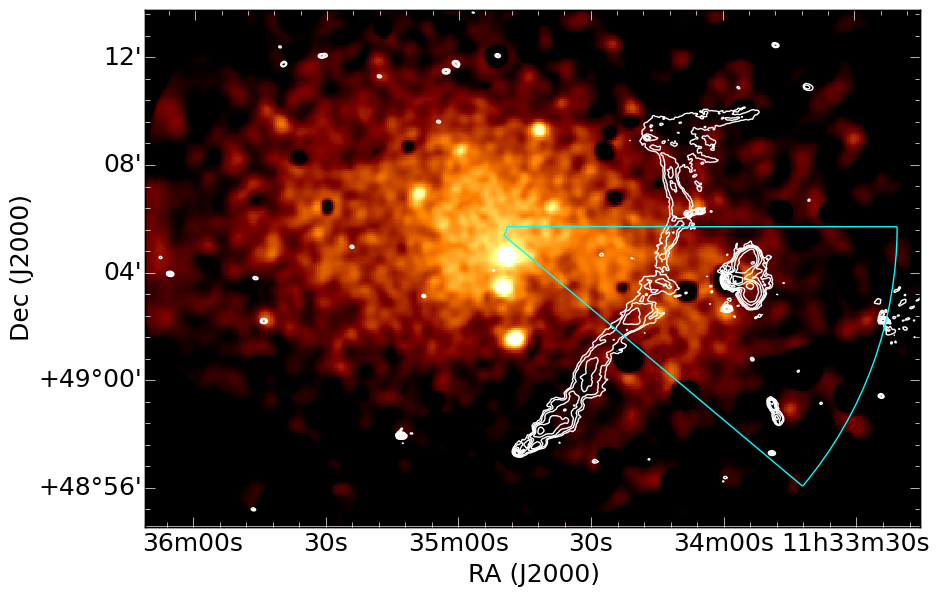}
\caption{XMM-Newton map of Abell 1314 with LOFAR high-resolution overlaid as white contours $[6, 12, 24, 48, 96, 192, 384, 768]~\times~\sigma$ where $\sigma = 180~\mu$Jy beam$^{-1}$. The cyan region is where the surface brightness is analysed along the length of the tail, starting from the cluster center and going to a radius of 600 kpc.  \label{xray}}
\end{figure}

\begin{figure*}
\begin{tabular}{cc}
\begin{minipage}{.5\hsize}
\begin{center}
\includegraphics[width=1.\hsize]{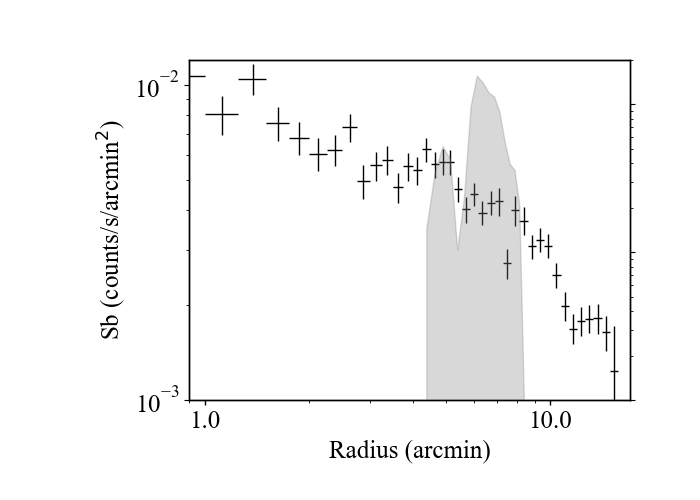}
\end{center}
\end{minipage}
\begin{minipage}{.5\hsize}
\begin{center}
\includegraphics[width=1.\hsize]{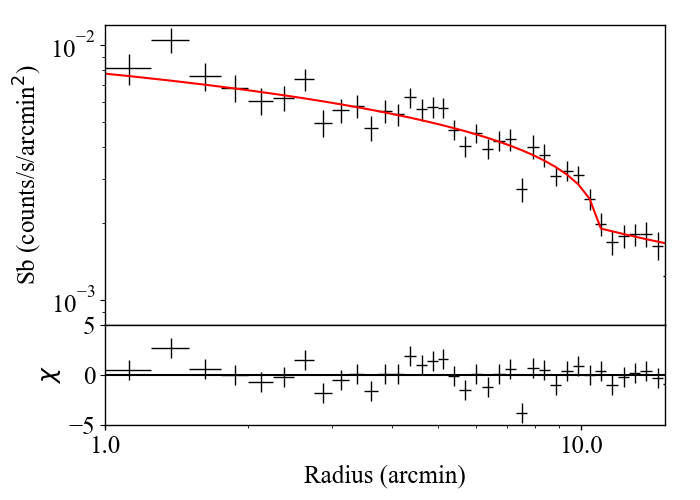}
\end{center}
\end{minipage}
\end{tabular}
\caption{\label{sb}
Left: \textit{XMM-Newton} 0.3 - 2.0 keV band surface brightness profile.
The gray shaded area represents the LOFAR radio brightness profile (in arbitrary units).
Right: Same as left but with the best-fitting broken power law model.
}
\end{figure*}

\subsubsection{X-ray results on Abell 1314 \label{a1314xray}}

Fig.~\ref{xray} shows the 0.3--2.0 keV X-ray band image of Abell 1314 obtained with \textit{XMM-Newton} together with LOFAR high-resolution radio contours. As is clear from Fig.~\ref{xray}, Abell 1314 shows an elongated and disturbed morphology indicating that it is still at a dynamically young state.\\

Using \textit{XMM-Newton} MOS data, we characterized the dynamical state of cluster and investigate the relationship between thermal (ICM) and non-thermal (radio) components. We derived the morphological parameters ($c_{100~\rm kpc}$ and $w_{500~\rm kpc}$) as $0.063\pm0.003$, $0.026\pm0.01$, respectively. The morphological parameters of Abell 1314 are in good agreement with those observed for merging clusters (see Fig.~\ref{wc-diagram}); however, we note that the values for Abell 1314 are computed from XMM data whereas the values of other clusters we use for comparison are computed from Chandra data. XMM has a larger point spread function (PSF), and this is taken into account in the errors on $w$ and $c$. \\

For the surface brightness profile, we used 0.3--2.0 keV energy range and fitted it with the PROFFIT software package \citep{2011A&A...526A..79E}. We extracted the surface brightness profile in the western sector with an opening angle of 40 degrees (covering the full region where the spectral index is seen to flatten). The resulting surface brightness profile is shown in the Fig.~\ref{sb}. The discontinuity in XMM-Newton surface brightness profile is clearly visible. The location of the discontinuity around \textit{r}$\sim$10\arcmin, is just outside of the BCG, IC\,712.
In order to characterize the discontinuity, we assume that the gas density follows two power-law profiles connecting at a discontinuity with a density jump. The density profile was projected onto the line-of-sight with the assumption of spherical symmetry.  All the parameters of the model were let free in the fit. 
The surface brightness profile was well-fitted with the above model (reduced $\chi^2$= 1.16 for 32 degrees of freedom).
The best-fitting broken power law model is shown in Fig. ~\ref{sb}. The compression factor and the location of the discontinuity are $\textit{C}=2.1\pm0.2$ and $r=10.8\pm0.3$ arcmin, respectively.\\

Since the discontinuity is located near the edge of the FOV of the MOS instrument, the determination of the instrumental background might be inaccurate. We investigate the impact of this systematic effect by changing the normalization of the instrumental background $\pm$20\%\footnote{We use a conservative, and likely overestimated, error value of 20\% and note that the results and 
conclusion do not change with or without the systematic error.}. The effect of the systematic error on the location of the discontinuity and the compression factor are smaller than or compatible with the statistical errors.
In the following discussion section, we used the error defined by $\displaystyle{\sigma\equiv\sqrt{\sigma_{\rm stat}^2+\sigma_{\rm syst}^2}}$, which leads $\textit{C}=2.1\pm0.4$ and $r=10.8\pm0.4$ arcmin, respectively.\\

Since Abell 1314 is not detected by Planck, we must estimate its mass from X-ray data. We estimate the mass of Abell 1314 from the spectrum of the \textit{XMM-Newton} observation. The 0.5-2.4 keV band flux of Abell 1314 is $F_x =6.13 \times 10^{-12}$ erg s$^1$ cm$^2$. The luminosity distance of Abell 1314 ($z\sim0.034$) is 150.3 Mpc therefore, the estimated X-ray luminosity is $L_x =1.7 \times 10^{43}$ erg s$^{-1}$ and the estimated ICM temperature is $kT \sim 1.7$ keV. Our result is in good agreement with the $L_x - T$ relation from \citet{2000ApJ...538...65X} (Fig.~1 in their paper). With this temperature we estimate the mass with the scaling relation in \citet{2011A&A...535A...4R}, which is based on 14 literature samples, and find $M_{500}\sim 0.68 \times 10^{14} M_{\odot}$. Our estimated mass is broadly consistent with the one from MCXC ($M_{500} = 0.4608 \times 10^{14}$ M$_{\odot}$; \citealt{2011A&A...534A.109P} )\footnote{There is a $\sim 30\%$ difference between our mass estimate and the mass estimate from \citealt{2011A&A...534A.109P}, which is within the scatter of the $L_x - M$ scaling relationship.}

\begin{figure*}
\centering
\includegraphics[width=\textwidth]{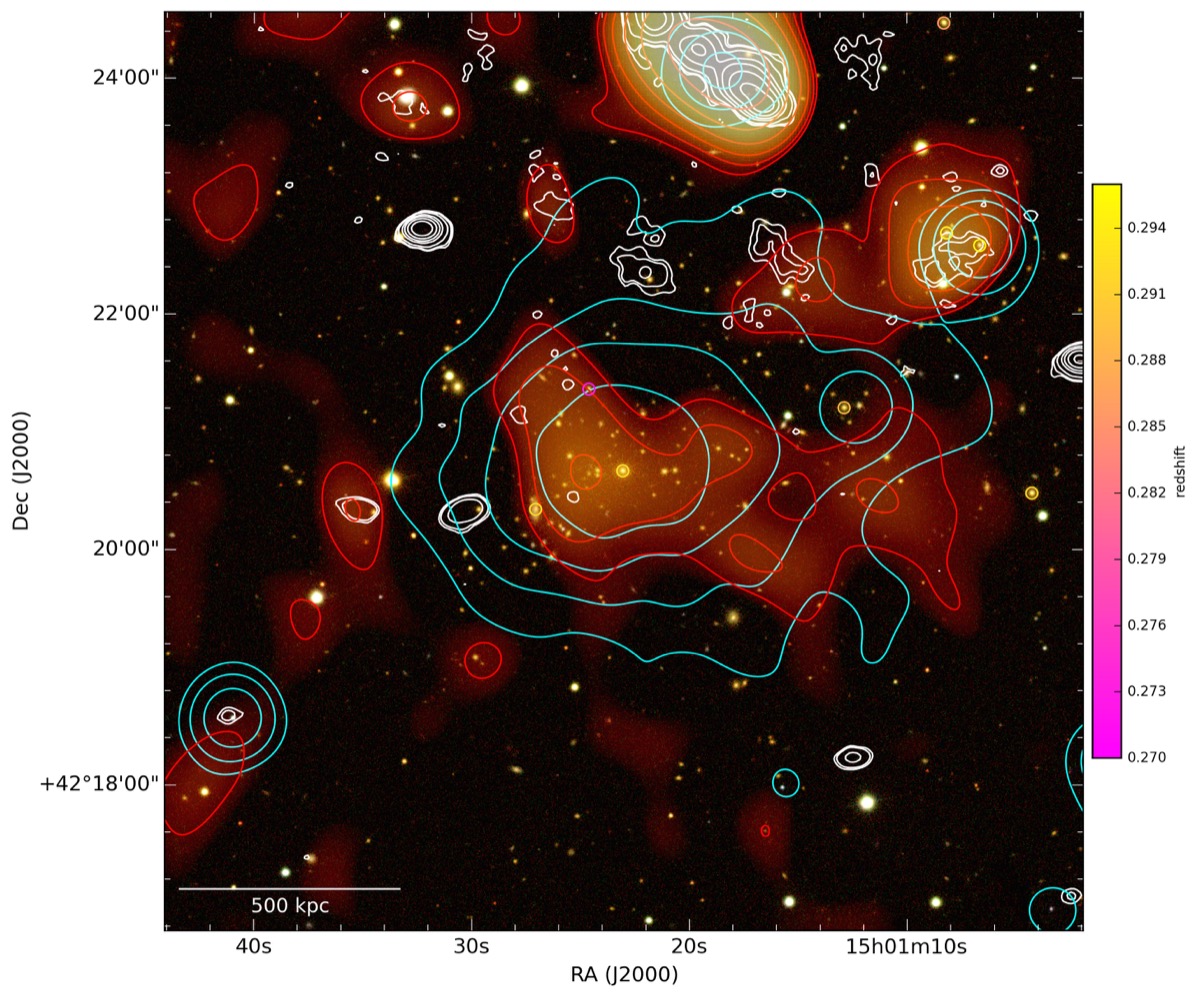}
\caption{SDSS {\it i,r,g} image of Z7215 with LOFAR radio emission overlaid as contours. LOFAR high-resolution (10 arcsec $\times$ 6 arcsec) emission is shown by white contours $[3,5,16,64,128,256,512]~\times~\sigma$ where $\sigma = 170~\mu$Jy beam$^{-1}$. LOFAR diffuse emission after compact source subtraction (with a resolution of 40 arcsec $\times$ 34 arcsec) is shown in red color and red contours where contours are $[2,4,8,16,32,64]~\times~\sigma$ and $\sigma=350~\mu$Jy beam$^{-1}$. Smoothed Chandra X-ray contours as also shown in cyan. Cluster galaxies with known spectroscopic redshifts are marked by colored rings, where the color represents redshift.  \label{Z7215opt}}
\end{figure*}

\subsection{Z7215: Example of post-merging system \label{Z7215res}}
An overlay of our LOFAR images of the cluster Z7215, in high- and low-resolution, can be seen in Fig.~\ref{Z7215opt}. This overlay includes smoothed X-ray contours from Chandra data. According to the Chandra map, the thermal ICM appears to be moderately disturbed and elongated. The cluster does not host a cool-core and looks moderately asymmetric. Because of the low counts and the position of the chips we cannot give reasonable estimates of $w$ and $c$ from the Chandra data alone. An \textit{XMM-Newton} observation of Z7215 is used instead to estimate values for the centroid shift and concentration parameters, as was done for Abell 1314, but these estimates have larger errors: $w= 0.096 \pm 0.06$ and $c= 0.052 \pm 0.09$. These values and the estimated $\beta$ parameter of $0.84 \pm 0.07$ indicate a disturbed morphology. The estimated core radius $R_c$ is $1.29 \pm 0.10$ arcmin. Considering the distance to this cluster (1 arcmin $\sim$ 260 kpc), the core radius is $\sim 350$ kpc, which is apparently on the larger side \citep[e.g.][]{1999ApJ...517..627M}.\\

The LOFAR image of Z7215 at high resolution does not show many radio sources associated with or located at the cluster center. A FR-II radio galaxy is seen north of the cluster, coincident with an X-ray source seen by Chandra, but this source is likely in the background since it overlaps galaxies at higher redshifts. After subtracting sources and re-imaging at a lower resolution, diffuse emission is visible above $3\sigma$ at the cluster center, where $\sigma = 450 ~\mu$ Jy beam$^{-1}$. This diffuse emission is coincident with the Chandra X-ray contours at the cluster center. However, this diffuse emission is not spherical in morphology, and instead appears to be elongated to the west, with a minor axis about half the size of the major axis (about 500 kpc in the N-S direction and 1 Mpc in the E-W direction). There is also a portion of diffuse emission detached from the center on the N-W edge, but it overlaps several potentially active galaxies, coincides with an X-ray source seen by Chandra, and is also more compact in its extent. We do not include this emission on the northwest edge, which is likely of AGN origin, as part of the central diffuse emission when measuring flux densities and calculating powers in the following sub-sections.\\

\subsubsection{Central diffuse radio emission \label{z7215res}}

We discover centrally located diffuse emission in Z7215 which is most likely a radio halo, although it is not spherically uniform in its shape and it is dimmer than expected from the $P-M$ correlation. The flux density of the central diffuse emission is measured in our compact-source-subtracted image made with a $uv$-taper of 30 arcsec (resolution of 43 arcsec $\times$ 33 arcsec). The flux density above $2\sigma$ contours where $\sigma = 450~\mu$Jy beam$^{-1}$ is $20.2 \pm 2.0 $ mJy. This translates to a power of $(5.91 \pm 0.60) \times 10^{24}~$W Hz$^{-1}$ at 144 MHz and $(3.07 \pm 0.31) \times 10^{23}~$W Hz$^{-1}$ at 1.4 GHz if extrapolated assuming a spectral index of $\alpha = -1.3$. This power of $\log_{10}$(P$_{1.4}$ $ /$ W Hz$^{-1}) = 23.49$ is slightly less than the expected power of a halo fitting on the correlation at 1.4 GHz (see Table~\ref{tab1}), but close to the powers of other radio halos at this cluster mass (see Fig.~\ref{corr}). A 610 MHz observation covering Z7215, as part of the GMRT Radio Halo Survey, was analysed by \citet{2008A&A...484..327V}, and they found an upper limit for diffuse radio emission at the cluster centre: $\log_{10}$(P$_{610}$ $ /$ W Hz$^{-1}) = 24.20$. Extrapolating this value to 1.4 GHz with a spectral index of $\alpha = -1.3$ gives $\log_{10}$(P$_{610}$ $ /$ W Hz$^{-1}) = 23.73$. This value is comparable but slightly larger than the power we extrapolate from our 144 MHz LOFAR observations. \\

To test the capability of our LOFAR observation to recover such faint extended emission, we inject a modeled radio halo with a radius $R_H$ determined by the correlations from \citet{2013ApJ...777..141C} ($R_{H} = 422$ kpc and $r_e$ = 162 kpc for a cluster with $M_{500} = 5.87\times10^{14}M_{\odot}$). We inject the mock radio halo in a relatively empty region of the cluster field (at RA, dec: 15h03m04.3s, +42d34m49.4s). The mock halo is initially assumed to have a power and central brightness, $I_0$, defined by the correlation, and we lower the injected flux until the recovered mock halo emission is comparable to the detected emission at the cluster center. We measure and compare the recovered flux density of the injected source to the flux density of the diffuse emission at the cluster centre above $2\sigma$ and determine at which injected flux value they are approximately equal. The flux density of the recovered halo is measured in our LOFAR low-resolution image\footnote{Made with a $uv$-taper of 30 arcsec and Briggs' robust parameter of 0.} above $2\sigma$ contours, and is found to be $19.3 \pm 1.9 ~$mJy. This recovered flux density of the mock halo is about equal, within error, to the central diffuse source. The injected flux value of this halo is 26.3 mJy, and therefore the recovered flux is $\sim 73\%$ of the injected flux. These findings demonstrate that our LOFAR observation is able to partially, but not fully, recover the faint extended emission of this radio halo. Therefore, we conclude that the true flux density of the radio halo in Z7215 is more reasonably approximated by the injected flux density value of the mock halo. We extrapolate this flux density to 1.4 GHz assuming a spectral index of $\alpha=-1.3$ and find a power of $3.99 \times 10^{23}$ W Hz$^{-1}$. We compare this power to the power of the detected emission above $2\sigma$ and to the powers of a sample of radio halos, as shown in Fig.~\ref{corr}. 

\section{Discussion}

\subsection{Abell 1319}
The three ICM clumps, A, B, and C, are separated by a distance of about $\sim$ 500 kpc. It is not likely that this system is post-merger, as we might then expect to see X-ray filaments or an overall elongated, disturbed ICM of the combined components. However, a deeper X-ray observation would be needed to exclude the presence of such filaments. Additionally, if this system was post-merger the ICM of Abell 1319-A would be expected to be more disturbed, with a higher $w$ value and a lower $c$ value. \\

Diffuse cluster-scale sources do not appear in our LOFAR images. We suggest that there are no radio relics or radio halos associated with this multi-component system. The lack of ICM radio sources lends further support to the hypothesis that this system is in its pre-merging phase since radio relics and radio halos are reliable tracers of mergers.

\subsection{Abell 1314 \label{disA1314}}

Although there is a central diffuse source in this cluster, its morphology, scale, and brightness distribution is unlike that of a cluster radio halo. The power of the central diffuse source lies an order of magnitude below the $P-M$ correlation, even at the low end of the mass range, suggesting that this source is not similar to a radio halo. Indeed, this cluster is considered to have a very low mass, and the $P-M$ correlation has not yet been tested for such low-mass merging clusters. Although Abell 1314 is highly disturbed (see Fig.~\ref{xray} and $w$, $c$ parameters in Table~\ref{tab1}), it is not so surprising that a halo is not detectable given the mass estimate we derive. It is possible that the ICM has been re-energized by merger turbulence, but a spherical, cluster-wide halo has not been generated.  \\

One may speculate whether this elongated patch of radio emission is the beginning, or making, of a radio halo, that will develop into a full-scale halo at a later stage. \citet{2013MNRAS.429.3564D} used simulations to show that radio halos are transient sources with an evolving spectrum, and identified three stages of radio / X-ray activity: infall, re-acceleration, and decay. It is unlikely that Abell 1314 is in the re-acceleration phase where a cluster-scale, spherically uniform halo is expected to form. Our deep observations should have revealed such a halo, however, we only detect faint ``arms'' of emission at the center of Abell 1314.\\

The central diffuse source, which likely has a steep-spectrum since it is not detected by GMRT observations at 235 or 610 MHz, (an estimate of the spectral index in Sec.~\ref{secspec} gives $\alpha \sim -1.3$) is similar to the irregular large-scale emission recently discovered in Abell 1931 by \citet{2018MNRAS.477.3461B}. In that paper, they report a remnant radio galaxy that is detected by LOFAR at 144 MHz. Another 144 MHz LOFAR observation of the galaxy group MaxBCG J199.31832+51.72503 showed an extended radio galaxy embedded in steep, remnant radio emission at the group's center \citep{2018MNRAS.474.5023S}. Here in Abell 1314 we may be witnessing a similar phenomena where the ``arms'' of emission are old, fading remnant lobes of a previous activity cycle from the central BCG AGN. However, there are multiple galaxies at the cluster redshift that reside within this diffuse radio source, hence the radio emission could be a superposition of multiple remnant lobes. A question remains as to whether this diffuse source is simply fading AGN emission, or if this fossil emission has been partially re-accelerated, and therefore slightly re-brightened, by merger turbulence. Without detailed spectral index maps this will be impossible to untangle.\\ 

Still, this is yet another example that LOFAR is able to detect old, faded, and faint large-scale radio sources. This fossil radio emission proves that relativistic electrons are filling the inner part of the cluster volume. As LOFAR reveals more and more fossil radio emission from remnant AGN, we may begin to see a clear connection between these sources and ICM cluster-scale sources such as radio halos \citep[e.g.][]{2018MNRAS.473.3536W, 2018MNRAS.477.3461B}.  \\

\begin{figure}
\centering
\includegraphics[width=0.49\textwidth]{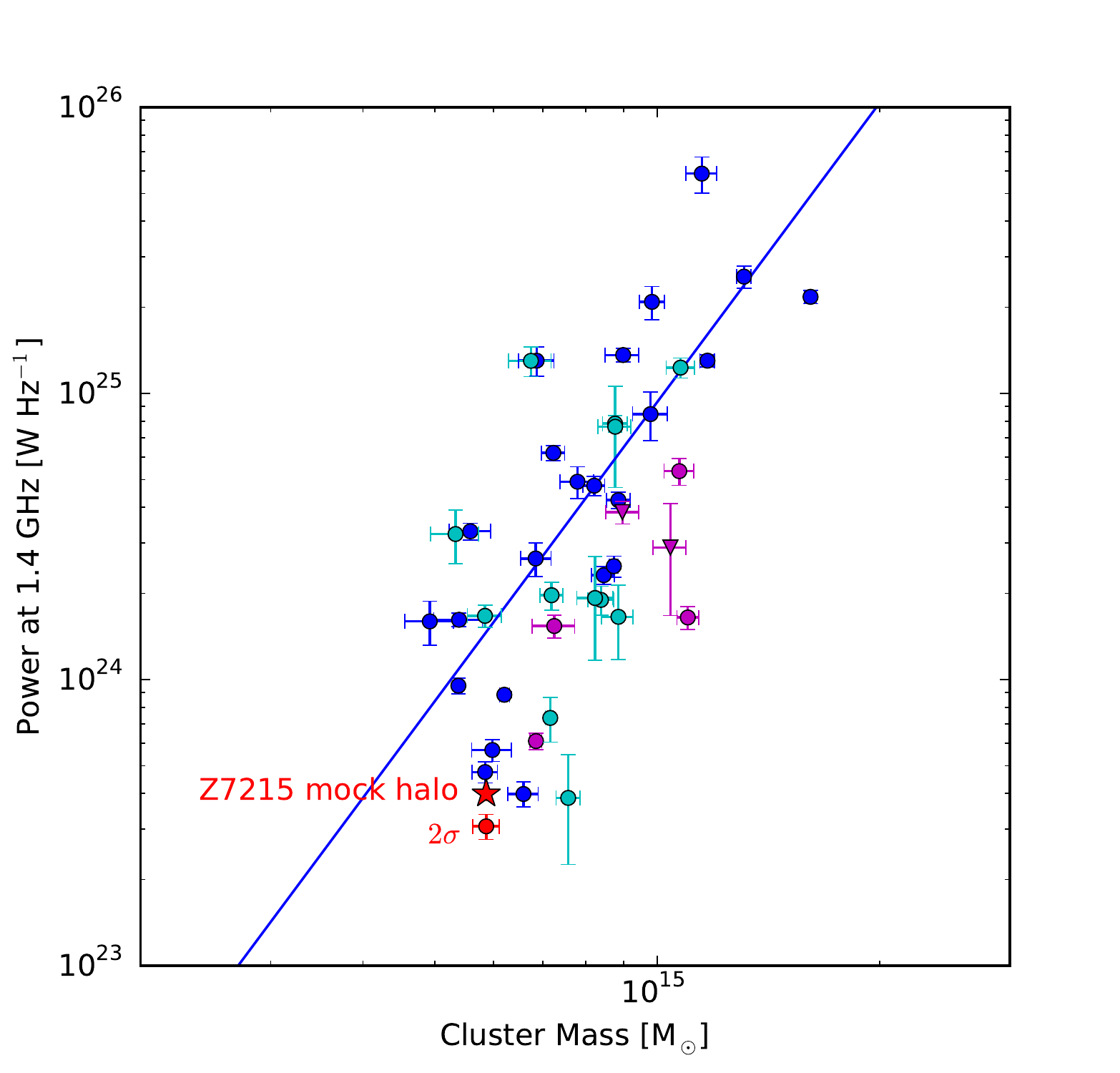}
\caption{A sample of radio halos plotted by their radio power at 1.4 GHz versus their cluster mass (M$_{500}$ -- as determined from Planck observations). The sample of halos and their correlation is reproduced from \citet{2016A&A...595A.116M}. Halos with flux measured at 1.4 GHz are marked by blue circles and their derived fit is shown as a blue line. Cyan circles represent halos with flux measured at frequencies other than 1.4 GHz. Magenta circles represent ultra-steep halos, and magenta triangles represent ultra-steep halos with flux measured at frequencies other than 1.4 GHz. The red point shows the power computed from the emission detected in our LOFAR observation for  Z7215. The red star is the power of our mock halo with a higher injected flux value. All halo powers include a k-correction with an averaged spectral index of $\alpha = -1.3$, as in \citet{2016A&A...595A.116M}. \label{corr}}
\end{figure}

\subsubsection*{IC~711} 

The extended radio emission produced by the galaxy IC~711 shows a spectral index that generally steepens from the head to the end of the tail (see Fig.~\ref{spix}), as is expected of typical head-tail radio galaxies. However, there is a portion of the tail, near the middle, that shows signs of flattening. This flattening likely occurs because the emission in this region is disturbed or compressed. We suggest that the disturbance induces re-acceleration which causes the electrons to emit at slightly higher energies in this region. In fact, the flattening occurs where there is a concentration of thermal ICM emission, in the same plane of the cluster merger. This flattening is similar to the gentle re-energization (GReET) phenomena reported by \citet{2017SciA....3E1634D}, where the spectral index of a WAT source is seen to increase where it is expected to decrease. \\

From a radial X-ray surface brightness profile of the cluster a small jump can be seen at the location of the radio tail of IC~711. A discontinuity is also seen in the eastward direction (not shown), but there are no significant radio sources on this side of the cluster. It is possible that bulk motions in the ICM distort the tail which flows perpendicular to the merger direction. The total shape of the tail has a slight arc, concave with respect to the cluster centre. In the high-resolution image, we see thin filamentary structures in the northern emission of the tail, which appear to be sweeping from east to west. This filamentary emission is even more suggestive of disturbance, such that this sweeping effect may have been produced by turbulence and shocks traveling from the cluster center to the outskirts. It is impossible to tell whether this northernmost emission has been re-accelerated, or compressed by a shock, because there is no radio emission in this region present in our GMRT images, and the XMM-Newton map is not sensitive enough to detect a jump in temperature or surface brightness in this region. The 240 MHz GMRT image produced by \citet{2017AJ....154..169S} may be helpful to estimate a spectral index in this region, but such an estimate could be unreliable since their observing frequency is close to our LOFAR observing frequency.

\subsection{Z7215}
The power of the diffuse source at the cluster center is below the $P-M$ correlation, but comparable to the powers of other radio halos at a similar cluster mass (see Fig.~\ref{corr}). It is likely that this diffuse source, elongated in the E-W direction and coincident with the thermal X-ray emission of the cluster, is a radio halo. There are no compact sources seen in our high-resolution LOFAR image that reside within the cluster center, indicating that it unlikely that the diffuse emission is of AGN origin. The flux density of the central diffuse source in Z7215 (contained within $2\sigma$ contours) is roughly equal to the flux density of the emission recovered after injecting a fake radio halo with a flux density of 26.3 mJy at 144 MHz.\\ 

Based on our findings, the radio halo in Z7215 is underluminous. One possibility for this could be that the halo is smaller than the typical halos fitting the correlation. Indeed the halo as seen by LOFAR appears shortened in the N-S direction, and we measure an effective radius $R_H= \sqrt{500 {\rm kpc} \times 250 {\rm kpc}} \approx 350$ kpc. Another possibility is that the halo is an ultra-steep spectrum radio halo (USSRH). USSRHs are predicted \citep{2010ApJ...721L..82C,2013MNRAS.429.3564D} and observed to be typically underluminous with respect to the correlation (see Fig.~\ref{corr}).
USSRHs are thought to form in less massive and less energetic merger events or at the beginning or fading evolutionary phase of halo formation \citep{2006MNRAS.369.1577C, 2008Natur.455..944B,2013MNRAS.429.3564D}. Since the actual halo radius is smaller than the radius assumed for the upper limit computed by \citet{2008A&A...484..327V}, the upper limit should be smaller, and the resulting spectral index should be steeper. Deeper observations around 330 MHz could help to constrain the spectral index.\\

This cluster was also on the edge of our LoTSS observation pointing, which may have affected the amount of emission that was detected and recovered. Indeed, only 73\% of the injected halo flux was recovered in our LOFAR image. Follow-up observations centered on this cluster are already planned and scheduled with LOFAR. Observations centered on the cluster may reveal more flux associated with the radio halo. Future LOFAR LBA observations may also be of interest in conjunction with the HBA observations so that the spectral index could be constrained. In conclusion we claim that this cluster is in its post-merging phase, where the system has begun to relax. The radio halo we discover is under-luminous with respect to the correlation. 

\section{Conclusions}

In Summary we find following results:
\begin{itemize}

\item We show that Abell 1319 is a multiple-component system and suggest that these three low-mass clusters, or galaxy groups, are about to merge. LOFAR radio observations reveal active BCG galaxies at the centers of all three ICMs. There are no diffuse cluster-scale ICM radio sources present in this system, such as radio relics or radio halos, which supports the notion that this system is pre-merging. \\

\item Abell 1314 is a low-mass cluster that shows a highly-disturbed thermal ICM. Irregularly-shaped diffuse radio emission present at the cluster center is likely to be remnant AGN lobe emission from previous activity cycles of the currently active BCG IC~712. We argue that this cluster is in a merging phase, although it is likely too low of mass to generate a detectable radio halo, even at a later evolutionary stage. The extended head-tail radio galaxy IC~711 shows signs of disturbance. A spectral index map reveals that the middle portion of the tail, where the X-ray brightness is highest along the tail, indicates possible local re-acceleration or compression. The northernmost part of the tail shows filamentary emission that appears to be sweeping westward, suggestive of disturbance.  \\

\item We discover a radio halo in the massive, merging cluster Z7215 and we argue that this cluster is in its post-merging phase. The halo has a non-spherical morphology with a major axis of about 1 Mpc and a minor axis of about 500 Mpc. The halo is dimmer than expected by the halo $P-M$ correlation but comparable to the powers of other halos at similar cluster masses. We suggest that this could be a smaller halo or an ultra-steep spectrum halo. Future follow-ups will distinguish between these possibilities. \\

\end{itemize}

Using LOFAR observations to add more clusters to this sample, specifically adding more examples of clusters which are in various stages of merging, would help to strengthen our understanding of how merging mechanisms are related to ICM radio sources.

\begin{acknowledgements}
LOFAR, the Low Frequency Array designed and constructed by ASTRON, has facilities owned by various parties (each with their own funding sources), and that are collectively operated by the International LOFAR Telescope (ILT) foundation under a joint scientific policy. The LOFAR software and dedicated reduction packages on https://github.com/apmechev/GRID\_LRT were deployed on the e-infrastructure by the LOFAR e-infragroup, consisting of J. B. R. Oonk (ASTRON \& Leiden Observatory), A. P. Mechev (Leiden Observatory) and T. Shimwell (Leiden Observatory) with support from N. Danezi (SURFsara) and C. Schrijvers (SURFsara). The data used in work was in part processed on the Dutch national e-infrastructure with the support of SURF Cooperative through grant e-infra 160022 \& 160152. We thank the staff of the GMRT who have made these observations  possible. GMRT is run by the National Centre for Radio Astrophysics of the Tata Institute of Fundamental Research. This research made use of the NASA/IPAC Extragalactic Database (NED), which is operated by the Jet Propulsion Laboratory, California Institute of Technology, under contract with the National Aeronautics and Space Administration. AB acknowledges support from the ERC-StG 714245 DRANOEL. RJvW and HJAR acknowledge support from the ERC Advanced Investigator programme NewClusters 321271 and RJvW acknowledges the VIDI research programme with project number 639.042.729, which is financed by the Netherlands Organisation for Scientific Research (NWO). HA acknowledges the support of NWO via a Veni grant. FdG is supported by the VENI research programme with project number 639.041.542, which is financed by the Netherlands Organisation for Scientific Research (NWO).
\end{acknowledgements}

\bibliographystyle{aa}
\bibliography{A1314.bib}

\begin{thebibliography}{70}
\expandafter\ifx\csname natexlab\endcsname\relax\def\natexlab#1{#1}\fi

\bibitem[{{Bernardi} {et~al.}(2016){Bernardi}, {Venturi}, {Cassano},
  {Dallacasa}, {Brunetti}, {Cuciti}, {Johnston-Hollitt}, {Oozeer}, {Parekh}, \&
  {Smirnov}}]{2016MNRAS.456.1259B}
{Bernardi}, G., {Venturi}, T., {Cassano}, R., {et~al.} 2016, \mnras, 456, 1259

\bibitem[{{B{\"o}hringer} {et~al.}(2010){B{\"o}hringer}, {Pratt}, {Arnaud},
  {Borgani}, {Croston}, {Ponman}, {Ameglio}, {Temple}, \& {Dolag}}]{Bo2010}
{B{\"o}hringer}, H., {Pratt}, G.~W., {Arnaud}, M., {et~al.} 2010, \aap, 514,
  A32

\bibitem[{{Bonafede} {et~al.}(2017){Bonafede}, {Cassano}, {Br{\"u}ggen},
  {Ogrean}, {Riseley}, {Cuciti}, {de Gasperin}, {Golovich}, {Kale}, {Venturi},
  {van Weeren}, {Wik}, \& {Wittman}}]{2017MNRAS.470.3465B}
{Bonafede}, A., {Cassano}, R., {Br{\"u}ggen}, M., {et~al.} 2017, \mnras, 470,
  3465

\bibitem[{{Bonafede} {et~al.}(2014){Bonafede}, {Intema}, {Br{\"u}ggen},
  {Russell}, {Ogrean}, {Basu}, {Sommer}, {van Weeren}, {Cassano}, {Fabian}, \&
  {R{\"o}ttgering}}]{2014MNRAS.444L..44B}
{Bonafede}, A., {Intema}, H.~T., {Br{\"u}ggen}, M., {et~al.} 2014, \mnras, 444,
  L44

\bibitem[{{Botteon} {et~al.}(2018){Botteon}, {Shimwell}, {Bonafede},
  {Dallacasa}, {Brunetti}, {Mandal}, {van Weeren}, {Br{\"u}ggen}, {Cassano},
  {de Gasperin}, {Hoang}, {Hoeft}, {R{\"o}ttgering}, {Savini}, {White},
  {Wilber}, \& {Venturi}}]{2018MNRAS.478..885B}
{Botteon}, A., {Shimwell}, T.~W., {Bonafede}, A., {et~al.} 2018, \mnras, 478,
  885

\bibitem[{{Brienza} {et~al.}(2017){Brienza}, {Godfrey}, {Morganti}, {Prandoni},
  {Harwood}, {Mahony}, {Hardcastle}, {Murgia}, {R{\"o}ttgering}, {Shimwell}, \&
  {Shulevski}}]{2017A&A...606A..98B}
{Brienza}, M., {Godfrey}, L., {Morganti}, R., {et~al.} 2017, \aap, 606, A98

\bibitem[{{Br{\"u}ggen} {et~al.}(2018){Br{\"u}ggen}, {Rafferty}, {Bonafede},
  {van Weeren}, {Shimwell}, {Intema}, {R{\"o}ttgering}, {Brunetti}, {Di
  Gennaro}, {Savini}, {Wilber}, {O'Sullivan}, {Ensslin}, {De Gasperin}, \&
  {Hoeft}}]{2018MNRAS.477.3461B}
{Br{\"u}ggen}, M., {Rafferty}, D., {Bonafede}, A., {et~al.} 2018, \mnras, 477,
  3461

\bibitem[{{Brunetti} {et~al.}(2008){Brunetti}, {Giacintucci}, {Cassano},
  {Lane}, {Dallacasa}, {Venturi}, {Kassim}, {Setti}, {Cotton}, \&
  {Markevitch}}]{2008Natur.455..944B}
{Brunetti}, G., {Giacintucci}, S., {Cassano}, R., {et~al.} 2008, \nat, 455, 944

\bibitem[{{Brunetti} \& {Jones}(2014)}]{2014IJMPD..2330007B}
{Brunetti}, G. \& {Jones}, T.~W. 2014, International Journal of Modern Physics
  D, 23, 1430007

\bibitem[{{Brunetti} {et~al.}(2001){Brunetti}, {Setti}, {Feretti}, \&
  {Giovannini}}]{2001MNRAS.320..365B}
{Brunetti}, G., {Setti}, G., {Feretti}, L., \& {Giovannini}, G. 2001, \mnras,
  320, 365

\bibitem[{{Brunetti} {et~al.}(2007){Brunetti}, {Venturi}, {Dallacasa},
  {Cassano}, {Dolag}, {Giacintucci}, \& {Setti}}]{2007ApJ...670L...5B}
{Brunetti}, G., {Venturi}, T., {Dallacasa}, D., {et~al.} 2007, \apjl, 670, L5

\bibitem[{{Carilli} \& {Taylor}(2002)}]{2002ARA&A..40..319C}
{Carilli}, C.~L. \& {Taylor}, G.~B. 2002, \araa, 40, 319

\bibitem[{{Cassano}(2010)}]{2010A&A...517A..10C}
{Cassano}, R. 2010, \aap, 517, A10

\bibitem[{{Cassano} \& {Brunetti}(2005)}]{2005MNRAS.357.1313C}
{Cassano}, R. \& {Brunetti}, G. 2005, \mnras, 357, 1313

\bibitem[{{Cassano} {et~al.}(2016){Cassano}, {Brunetti}, {Giocoli}, \&
  {Ettori}}]{2016A&A...593A..81C}
{Cassano}, R., {Brunetti}, G., {Giocoli}, C., \& {Ettori}, S. 2016, \aap, 593,
  A81

\bibitem[{{Cassano} {et~al.}(2012){Cassano}, {Brunetti}, {Norris},
  {R{\"o}ttgering}, {Johnston-Hollitt}, \& {Trasatti}}]{2012A&A...548A.100C}
{Cassano}, R., {Brunetti}, G., {Norris}, R.~P., {et~al.} 2012, \aap, 548, A100

\bibitem[{{Cassano} {et~al.}(2006){Cassano}, {Brunetti}, \&
  {Setti}}]{2006MNRAS.369.1577C}
{Cassano}, R., {Brunetti}, G., \& {Setti}, G. 2006, \mnras, 369, 1577

\bibitem[{{Cassano} {et~al.}(2007){Cassano}, {Brunetti}, {Setti}, {Govoni}, \&
  {Dolag}}]{2007MNRAS.378.1565C}
{Cassano}, R., {Brunetti}, G., {Setti}, G., {Govoni}, F., \& {Dolag}, K. 2007,
  \mnras, 378, 1565

\bibitem[{{Cassano} {et~al.}(2013){Cassano}, {Ettori}, {Brunetti},
  {Giacintucci}, {Pratt}, {Venturi}, {Kale}, {Dolag}, \&
  {Markevitch}}]{2013ApJ...777..141C}
{Cassano}, R., {Ettori}, S., {Brunetti}, G., {et~al.} 2013, \apj, 777, 141

\bibitem[{{Cassano} {et~al.}(2010){Cassano}, {Ettori}, {Giacintucci},
  {Brunetti}, {Markevitch}, {Venturi}, \& {Gitti}}]{2010ApJ...721L..82C}
{Cassano}, R., {Ettori}, S., {Giacintucci}, S., {et~al.} 2010, \apjl, 721, L82

\bibitem[{{Cornwell} {et~al.}(2005){Cornwell}, {Golap}, \&
  {Bhatnagar}}]{2005ASPC..347...86C}
{Cornwell}, T.~J., {Golap}, K., \& {Bhatnagar}, S. 2005, in Astronomical
  Society of the Pacific Conference Series, Vol. 347, Astronomical Data
  Analysis Software and Systems XIV, ed. P.~{Shopbell}, M.~{Britton}, \&
  R.~{Ebert}, 86

\bibitem[{{Cuciti} {et~al.}(2018){Cuciti}, {Brunetti}, {van Weeren},
  {Bonafede}, {Dallacasa}, {Cassano}, {Venturi}, \&
  {Kale}}]{2018A&A...609A..61C}
{Cuciti}, V., {Brunetti}, G., {van Weeren}, R., {et~al.} 2018, \aap, 609, A61

\bibitem[{{Cuciti} {et~al.}(2015){Cuciti}, {Cassano}, {Brunetti}, {Dallacasa},
  {Kale}, {Ettori}, \& {Venturi}}]{2015A&A...580A..97C}
{Cuciti}, V., {Cassano}, R., {Brunetti}, G., {et~al.} 2015, \aap, 580, A97

\bibitem[{{Dallacasa} {et~al.}(2009){Dallacasa}, {Brunetti}, {Giacintucci},
  {Cassano}, {Venturi}, {Macario}, {Kassim}, {Lane}, \&
  {Setti}}]{2009ApJ...699.1288D}
{Dallacasa}, D., {Brunetti}, G., {Giacintucci}, S., {et~al.} 2009, \apj, 699,
  1288

\bibitem[{{de Gasperin} {et~al.}(2018){de Gasperin}, {Dijkema}, {Drabent}, \&
  {et~al.}}]{deGasperin2018a}
{de Gasperin}, F., {Dijkema}, T.~J., {Drabent}, A., \& {et~al.} 2018, A{\&}A

\bibitem[{{de Gasperin} {et~al.}(2017){de Gasperin}, {Intema}, {Shimwell},
  {Brunetti}, {Br{\"u}ggen}, {En{\ss}lin}, {van Weeren}, {Bonafede}, \&
  {R{\"o}ttgering}}]{2017SciA....3E1634D}
{de Gasperin}, F., {Intema}, H.~T., {Shimwell}, T.~W., {et~al.} 2017, Science
  Advances, 3, e1701634

\bibitem[{{Donnert} {et~al.}(2013){Donnert}, {Dolag}, {Brunetti}, \&
  {Cassano}}]{2013MNRAS.429.3564D}
{Donnert}, J., {Dolag}, K., {Brunetti}, G., \& {Cassano}, R. 2013, \mnras, 429,
  3564

\bibitem[{{Eckert} {et~al.}(2011){Eckert}, {Molendi}, \&
  {Paltani}}]{2011A&A...526A..79E}
{Eckert}, D., {Molendi}, S., \& {Paltani}, S. 2011, \aap, 526, A79

\bibitem[{Feretti {et~al.}(2012)Feretti, Giovannini, Govoni, \&
  Murgia}]{feretti2012}
Feretti, L., Giovannini, G., Govoni, F., \& Murgia, M. 2012, The Astronomy and
  Astrophysics Review, 20, 1

\bibitem[{{Giacintucci} {et~al.}(2009){Giacintucci}, {Venturi}, {Cassano},
  {Dallacasa}, \& {Brunetti}}]{2009ApJ...704L..54G}
{Giacintucci}, S., {Venturi}, T., {Cassano}, R., {Dallacasa}, D., \&
  {Brunetti}, G. 2009, \apjl, 704, L54

\bibitem[{{Giovannini} {et~al.}(2011){Giovannini}, {Feretti}, {Girardi},
  {Govoni}, {Murgia}, {Vacca}, \& {Bagchi}}]{2011A&A...530L...5G}
{Giovannini}, G., {Feretti}, L., {Girardi}, M., {et~al.} 2011, \aap, 530, L5

\bibitem[{{Golovich} {et~al.}(2018){Golovich}, {Dawson}, {Wittman}, {van
  Weeren}, {Andrade-Santos}, {Jee}, {Benson}, {de Gasperin}, {Venturi},
  {Bonafede}, {Sobral}, {Ogrean}, {Lemaux}, {Brada{\v c}}, {Br{\"u}ggen}, \&
  {Peter}}]{2018arXiv180610619G}
{Golovich}, N., {Dawson}, W.~A., {Wittman}, D.~M., {et~al.} 2018, ArXiv
  e-prints [\eprint[arXiv]{1806.10619}]

\bibitem[{{Hales} {et~al.}(2007){Hales}, {Riley}, {Waldram}, {Warner}, \&
  {Baldwin}}]{2007MNRAS.382.1639H}
{Hales}, S.~E.~G., {Riley}, J.~M., {Waldram}, E.~M., {Warner}, P.~J., \&
  {Baldwin}, J.~E. 2007, \mnras, 382, 1639

\bibitem[{{Intema} {et~al.}(2017){Intema}, {Jagannathan}, {Mooley}, \&
  {Frail}}]{2017A&A...598A..78I}
{Intema}, H.~T., {Jagannathan}, P., {Mooley}, K.~P., \& {Frail}, D.~A. 2017,
  \aap, 598, A78

\bibitem[{{Kale} {et~al.}(2015){Kale}, {Venturi}, {Giacintucci}, {Dallacasa},
  {Cassano}, {Brunetti}, {Cuciti}, {Macario}, \&
  {Athreya}}]{2015A&A...579A..92K}
{Kale}, R., {Venturi}, T., {Giacintucci}, S., {et~al.} 2015, \aap, 579, A92

\bibitem[{{Kale} {et~al.}(2013){Kale}, {Venturi}, {Giacintucci}, {Dallacasa},
  {Cassano}, {Brunetti}, {Macario}, \& {Athreya}}]{2013A&A...557A..99K}
{Kale}, R., {Venturi}, T., {Giacintucci}, S., {et~al.} 2013, \aap, 557, A99

\bibitem[{{Kravtsov} \& {Borgani}(2012)}]{2012ARA&A..50..353K}
{Kravtsov}, A.~V. \& {Borgani}, S. 2012, \araa, 50, 353

\bibitem[{{Martinez Aviles} {et~al.}(2016){Martinez Aviles}, {Ferrari},
  {Johnston-Hollitt}, {Pratley}, {Macario}, {Venturi}, {Brunetti}, {Cassano},
  {Dallacasa}, {Intema}, {Giacintucci}, {Hurier}, {Aghanim}, {Douspis}, \&
  {Langer}}]{2016A&A...595A.116M}
{Martinez Aviles}, G., {Ferrari}, C., {Johnston-Hollitt}, M., {et~al.} 2016,
  \aap, 595, A116

\bibitem[{{McMullin} {et~al.}(2007){McMullin}, {Waters}, {Schiebel}, {Young},
  \& {Golap}}]{2007ASPC..376..127M}
{McMullin}, J.~P., {Waters}, B., {Schiebel}, D., {Young}, W., \& {Golap}, K.
  2007, in Astronomical Society of the Pacific Conference Series, Vol. 376,
  Astronomical Data Analysis Software and Systems XVI, ed. R.~A. {Shaw},
  F.~{Hill}, \& D.~J. {Bell}, 127

\bibitem[{{Miley}(1980)}]{1980ARA&A..18..165M}
{Miley}, G. 1980, \araa, 18, 165

\bibitem[{{Mohr} {et~al.}(1999){Mohr}, {Mathiesen}, \&
  {Evrard}}]{1999ApJ...517..627M}
{Mohr}, J.~J., {Mathiesen}, B., \& {Evrard}, A.~E. 1999, \apj, 517, 627

\bibitem[{{Molendi} \& {Pizzolato}(2001)}]{2001ApJ...560..194M}
{Molendi}, S. \& {Pizzolato}, F. 2001, \apj, 560, 194

\bibitem[{{Murgia} {et~al.}(2010){Murgia}, {Govoni}, {Feretti}, \&
  {Giovannini}}]{2010A&A...509A..86M}
{Murgia}, M., {Govoni}, F., {Feretti}, L., \& {Giovannini}, G. 2010, \aap, 509,
  A86

\bibitem[{{Murgia} {et~al.}(2009){Murgia}, {Govoni}, {Markevitch}, {Feretti},
  {Giovannini}, {Taylor}, \& {Carretti}}]{2009A&A...499..679M}
{Murgia}, M., {Govoni}, F., {Markevitch}, M., {et~al.} 2009, \aap, 499, 679

\bibitem[{{Offringa} {et~al.}(2014){Offringa}, {McKinley}, {Hurley-Walker},
  {Briggs}, {Wayth}, {Kaplan}, {Bell}, {Feng}, {Neben}, {Hughes}, {Rhee},
  {Murphy}, {Bhat}, {Bernardi}, {Bowman}, {Cappallo}, {Corey}, {Deshpande},
  {Emrich}, {Ewall-Wice}, {Gaensler}, {Goeke}, {Greenhill}, {Hazelton},
  {Hindson}, {Johnston-Hollitt}, {Jacobs}, {Kasper}, {Kratzenberg}, {Lenc},
  {Lonsdale}, {Lynch}, {McWhirter}, {Mitchell}, {Morales}, {Morgan},
  {Kudryavtseva}, {Oberoi}, {Ord}, {Pindor}, {Procopio}, {Prabu}, {Riding},
  {Roshi}, {Shankar}, {Srivani}, {Subrahmanyan}, {Tingay}, {Waterson},
  {Webster}, {Whitney}, {Williams}, \& {Williams}}]{2014MNRAS.444..606O}
{Offringa}, A.~R., {McKinley}, B., {Hurley-Walker}, N., {et~al.} 2014, \mnras,
  444, 606

\bibitem[{{Petrosian}(2001)}]{2001ApJ...557..560P}
{Petrosian}, V. 2001, \apj, 557, 560

\bibitem[{{Piffaretti} {et~al.}(2011){Piffaretti}, {Arnaud}, {Pratt},
  {Pointecouteau}, \& {Melin}}]{2011A&A...534A.109P}
{Piffaretti}, R., {Arnaud}, M., {Pratt}, G.~W., {Pointecouteau}, E., \&
  {Melin}, J.-B. 2011, \aap, 534, A109

\bibitem[{{Planck Collaboration} {et~al.}(2014){Planck Collaboration}, {Ade},
  {Aghanim}, {Armitage-Caplan}, {Arnaud}, {Ashdown}, {Atrio-Barandela},
  {Aumont}, {Aussel}, {Baccigalupi}, \& et~al.}]{2014A&A...571A..29P}
{Planck Collaboration}, {Ade}, P.~A.~R., {Aghanim}, N., {et~al.} 2014, \aap,
  571, A29

\bibitem[{{Reichert} {et~al.}(2011){Reichert}, {B{\"o}hringer}, {Fassbender},
  \& {M{\"u}hlegger}}]{2011A&A...535A...4R}
{Reichert}, A., {B{\"o}hringer}, H., {Fassbender}, R., \& {M{\"u}hlegger}, M.
  2011, \aap, 535, A4

\bibitem[{{Santos} {et~al.}(2008){Santos}, {Rosati}, {Tozzi}, {B{\"o}hringer},
  {Ettori}, \& {Bignamini}}]{Santos2008}
{Santos}, J.~S., {Rosati}, P., {Tozzi}, P., {et~al.} 2008, \aap, 483, 35

\bibitem[{{Savini} {et~al.}(2018{\natexlab{a}}){Savini}, {Bonafede},
  {Br{\"u}ggen}, \& {et~al.}}]{Savini2018c}
{Savini}, F., {Bonafede}, A., {Br{\"u}ggen}, M., \& {et~al.}
  2018{\natexlab{a}}, A{\&}A

\bibitem[{{Savini} {et~al.}(2018{\natexlab{b}}){Savini}, {Bonafede},
  {Br{\"u}ggen}, {Wilber}, {Harwood}, {Murgia}, {Shimwell}, {Rafferty},
  {Shulevski}, {Brienza}, {Hardcastle}, {Morganti}, {R{\"o}ttgering}, {Clarke},
  {de Gasperin}, {van Weeren}, {Best}, {Botteon}, {Brunetti}, \&
  {Cassano}}]{2018MNRAS.474.5023S}
{Savini}, F., {Bonafede}, A., {Br{\"u}ggen}, M., {et~al.} 2018{\natexlab{b}},
  \mnras, 474, 5023

\bibitem[{{Scaife} \& {Heald}(2012)}]{2012MNRAS.423L..30S}
{Scaife}, A.~M.~M. \& {Heald}, G.~H. 2012, \mnras, 423, L30

\bibitem[{{Sebastian} {et~al.}(2017){Sebastian}, {Lal}, \& {Pramesh
  Rao}}]{2017AJ....154..169S}
{Sebastian}, B., {Lal}, D.~V., \& {Pramesh Rao}, A. 2017, \aj, 154, 169

\bibitem[{{Shimwell} {et~al.}(2016{\natexlab{a}}){Shimwell}, {Luckin},
  {Br{\"u}ggen}, {Brunetti}, {Intema}, {Owers}, {R{\"o}ttgering}, {Stroe}, {van
  Weeren}, {Williams}, {Cassano}, {de Gasperin}, {Heald}, {Hoang},
  {Hardcastle}, {Sridhar}, {Sabater}, {Best}, {Bonafede}, {Chy{\.z}y},
  {En{\ss}lin}, {Ferrari}, {Haverkorn}, {Hoeft}, {Horellou}, {McKean},
  {Morabito}, {Orr{\`u}}, {Pizzo}, {Retana-Montenegro}, \&
  {White}}]{2016MNRAS.459..277S}
{Shimwell}, T.~W., {Luckin}, J., {Br{\"u}ggen}, M., {et~al.}
  2016{\natexlab{a}}, \mnras, 459, 277

\bibitem[{{Shimwell} {et~al.}(2016{\natexlab{b}}){Shimwell}, {Luckin},
  {Br{\"u}ggen}, {Brunetti}, {Intema}, {Owers}, {R{\"o}ttgering}, {Stroe}, {van
  Weeren}, {Williams}, {Cassano}, {de Gasperin}, {Heald}, {Hoang},
  {Hardcastle}, {Sridhar}, {Sabater}, {Best}, {Bonafede}, {Chy{\.z}y},
  {En{\ss}lin}, {Ferrari}, {Haverkorn}, {Hoeft}, {Horellou}, {McKean},
  {Morabito}, {Orr{\`u}}, {Pizzo}, {Retana-Montenegro}, \&
  {White}}]{shimwell2016}
{Shimwell}, T.~W., {Luckin}, J., {Br{\"u}ggen}, M., {et~al.}
  2016{\natexlab{b}}, \mnras, 459, 277

\bibitem[{{Shimwell} {et~al.}(2017){Shimwell}, {R{\"o}ttgering}, {Best},
  {Williams}, {Dijkema}, {de Gasperin}, {Hardcastle}, {Heald}, {Hoang},
  {Horneffer}, {Intema}, {Mahony}, {Mandal}, {Mechev}, {Morabito}, {Oonk},
  {Rafferty}, {Retana-Montenegro}, {Sabater}, {Tasse}, {van Weeren},
  {Br{\"u}ggen}, {Brunetti}, {Chy{\.z}y}, {Conway}, {Haverkorn}, {Jackson},
  {Jarvis}, {McKean}, {Miley}, {Morganti}, {White}, {Wise}, {van Bemmel},
  {Beck}, {Brienza}, {Bonafede}, {Calistro Rivera}, {Cassano}, {Clarke},
  {Cseh}, {Deller}, {Drabent}, {van Driel}, {Engels}, {Falcke}, {Ferrari},
  {Fr{\"o}hlich}, {Garrett}, {Harwood}, {Heesen}, {Hoeft}, {Horellou},
  {Israel}, {Kapi{\'n}ska}, {Kunert-Bajraszewska}, {McKay}, {Mohan},
  {Orr{\'u}}, {Pizzo}, {Prandoni}, {Schwarz}, {Shulevski}, {Sipior}, {Smith},
  {Sridhar}, {Steinmetz}, {Stroe}, {Varenius}, {van der Werf}, {Zensus}, \&
  {Zwart}}]{2017A&A...598A.104S}
{Shimwell}, T.~W., {R{\"o}ttgering}, H.~J.~A., {Best}, P.~N., {et~al.} 2017,
  \aap, 598, A104

\bibitem[{{Srivastava} \& {Singal}(2016)}]{2016arXiv161007783S}
{Srivastava}, S. \& {Singal}, A.~K. 2016, ArXiv e-prints
  [\eprint[arXiv]{1610.07783}]

\bibitem[{{Vallee} \& {Wilson}(1976)}]{1976Natur.259..451V}
{Vallee}, J.~P. \& {Wilson}, A.~S. 1976, \nat, 259, 451

\bibitem[{{van Haarlem} {et~al.}(2013){van Haarlem}, {Wise}, {Gunst}, {Heald},
  {McKean}, {Hessels}, {de Bruyn}, {Nijboer}, {Swinbank}, {Fallows},
  {Brentjens}, {Nelles}, {Beck}, {Falcke}, {Fender}, {H{\"o}randel},
  {Koopmans}, {Mann}, {Miley}, {R{\"o}ttgering}, {Stappers}, {Wijers},
  {Zaroubi}, {van den Akker}, {Alexov}, {Anderson}, {Anderson}, {van Ardenne},
  {Arts}, {Asgekar}, {Avruch}, {Batejat}, {B{\"a}hren}, {Bell}, {Bell}, {van
  Bemmel}, {Bennema}, {Bentum}, {Bernardi}, {Best}, {B{\^i}rzan}, {Bonafede},
  {Boonstra}, {Braun}, {Bregman}, {Breitling}, {van de Brink}, {Broderick},
  {Broekema}, {Brouw}, {Br{\"u}ggen}, {Butcher}, {van Cappellen}, {Ciardi},
  {Coenen}, {Conway}, {Coolen}, {Corstanje}, {Damstra}, {Davies}, {Deller},
  {Dettmar}, {van Diepen}, {Dijkstra}, {Donker}, {Doorduin}, {Dromer}, {Drost},
  {van Duin}, {Eisl{\"o}ffel}, {van Enst}, {Ferrari}, {Frieswijk}, {Gankema},
  {Garrett}, {de Gasperin}, {Gerbers}, {de Geus}, {Grie{\ss}meier}, {Grit},
  {Gruppen}, {Hamaker}, {Hassall}, {Hoeft}, {Holties}, {Horneffer}, {van der
  Horst}, {van Houwelingen}, {Huijgen}, {Iacobelli}, {Intema}, {Jackson},
  {Jelic}, {de Jong}, {Juette}, {Kant}, {Karastergiou}, {Koers}, {Kollen},
  {Kondratiev}, {Kooistra}, {Koopman}, {Koster}, {Kuniyoshi}, {Kramer},
  {Kuper}, {Lambropoulos}, {Law}, {van Leeuwen}, {Lemaitre}, {Loose}, {Maat},
  {Macario}, {Markoff}, {Masters}, {McFadden}, {McKay-Bukowski}, {Meijering},
  {Meulman}, {Mevius}, {Middelberg}, {Millenaar}, {Miller-Jones}, {Mohan},
  {Mol}, {Morawietz}, {Morganti}, {Mulcahy}, {Mulder}, {Munk}, {Nieuwenhuis},
  {van Nieuwpoort}, {Noordam}, {Norden}, {Noutsos}, {Offringa}, {Olofsson},
  {Omar}, {Orr{\'u}}, {Overeem}, {Paas}, {Pandey-Pommier}, {Pandey}, {Pizzo},
  {Polatidis}, {Rafferty}, {Rawlings}, {Reich}, {de Reijer}, {Reitsma},
  {Renting}, {Riemers}, {Rol}, {Romein}, {Roosjen}, {Ruiter}, {Scaife}, {van
  der Schaaf}, {Scheers}, {Schellart}, {Schoenmakers}, {Schoonderbeek},
  {Serylak}, {Shulevski}, {Sluman}, {Smirnov}, {Sobey}, {Spreeuw}, {Steinmetz},
  {Sterks}, {Stiepel}, {Stuurwold}, {Tagger}, {Tang}, {Tasse}, {Thomas},
  {Thoudam}, {Toribio}, {van der Tol}, {Usov}, {van Veelen}, {van der Veen},
  {ter Veen}, {Verbiest}, {Vermeulen}, {Vermaas}, {Vocks}, {Vogt}, {de Vos},
  {van der Wal}, {van Weeren}, {Weggemans}, {Weltevrede}, {White}, {Wijnholds},
  {Wilhelmsson}, {Wucknitz}, {Yatawatta}, {Zarka}, {Zensus}, \& {van
  Zwieten}}]{vanHaarlem2013}
{van Haarlem}, M.~P., {Wise}, M.~W., {Gunst}, A.~W., {et~al.} 2013, \aap, 556,
  A2

\bibitem[{{van Weeren} {et~al.}(2017){van Weeren}, {Andrade-Santos}, {Dawson},
  {Golovich}, {Lal}, {Kang}, {Ryu}, {Br{\`i}ggen}, {Ogrean}, {Forman}, {Jones},
  {Placco}, {Santucci}, {Wittman}, {Jee}, {Kraft}, {Sobral}, {Stroe}, \&
  {Fogarty}}]{2017NatAs...1E...5V}
{van Weeren}, R.~J., {Andrade-Santos}, F., {Dawson}, W.~A., {et~al.} 2017,
  Nature Astronomy, 1, 0005

\bibitem[{{van Weeren} {et~al.}(2016){van Weeren}, {Williams}, {Hardcastle},
  {Shimwell}, {Rafferty}, {Sabater}, {Heald}, {Sridhar}, {Dijkema}, {Brunetti},
  {Br{\"u}ggen}, {Andrade-Santos}, {Ogrean}, {R{\"o}ttgering}, {Dawson},
  {Forman}, {de Gasperin}, {Jones}, {Miley}, {Rudnick}, {Sarazin}, {Bonafede},
  {Best}, {B{\^i}rzan}, {Cassano}, {Chy{\.z}y}, {Croston}, {En{\ss}lin},
  {Ferrari}, {Hoeft}, {Horellou}, {Jarvis}, {Kraft}, {Mevius}, {Intema},
  {Murray}, {Orr{\'u}}, {Pizzo}, {Simionescu}, {Stroe}, {van der Tol}, \&
  {White}}]{vanWeeren16}
{van Weeren}, R.~J., {Williams}, W.~L., {Hardcastle}, M.~J., {et~al.} 2016,
  \apjs, 223, 2

\bibitem[{{Venturi} {et~al.}(2007){Venturi}, {Giacintucci}, {Brunetti},
  {Cassano}, {Bardelli}, {Dallacasa}, \& {Setti}}]{2007A&A...463..937V}
{Venturi}, T., {Giacintucci}, S., {Brunetti}, G., {et~al.} 2007, \aap, 463, 937

\bibitem[{{Venturi} {et~al.}(2008){Venturi}, {Giacintucci}, {Dallacasa},
  {Cassano}, {Brunetti}, {Bardelli}, \& {Setti}}]{2008A&A...484..327V}
{Venturi}, T., {Giacintucci}, S., {Dallacasa}, D., {et~al.} 2008, \aap, 484,
  327

\bibitem[{{Vikhlinin} {et~al.}(2005){Vikhlinin}, {Markevitch}, {Murray},
  {Jones}, {Forman}, \& {Van Speybroeck}}]{Vikhlinin2005}
{Vikhlinin}, A., {Markevitch}, M., {Murray}, S.~S., {et~al.} 2005, \apj, 628,
  655

\bibitem[{{Webber}(1974)}]{1974PASP...86..223W}
{Webber}, J.~C. 1974, \pasp, 86, 223

\bibitem[{{Wilber} {et~al.}(2018{\natexlab{a}}){Wilber}, {Br{\"u}ggen},
  {Bonafede}, {Rafferty}, {Savini}, {Shimwell}, {van Weeren}, {Botteon},
  {Cassano}, {Brunetti}, {De Gasperin}, {Wittor}, {Hoeft}, \&
  {Birzan}}]{2018MNRAS.476.3415W}
{Wilber}, A., {Br{\"u}ggen}, M., {Bonafede}, A., {et~al.} 2018{\natexlab{a}},
  \mnras, 476, 3415

\bibitem[{{Wilber} {et~al.}(2018{\natexlab{b}}){Wilber}, {Br{\"u}ggen},
  {Bonafede}, {Savini}, {Shimwell}, {van Weeren}, {Rafferty}, {Mechev},
  {Intema}, {Andrade-Santos}, {Clarke}, {Mahony}, {Morganti}, {Prandoni},
  {Brunetti}, {R{\"o}ttgering}, {Mandal}, {de Gasperin}, \&
  {Hoeft}}]{2018MNRAS.473.3536W}
{Wilber}, A., {Br{\"u}ggen}, M., {Bonafede}, A., {et~al.} 2018{\natexlab{b}},
  \mnras, 473, 3536

\bibitem[{{Williams} {et~al.}(2016){Williams}, {van Weeren}, {R{\"o}ttgering},
  {Best}, {Dijkema}, {de Gasperin}, {Hardcastle}, {Heald}, {Prandoni},
  {Sabater}, {Shimwell}, {Tasse}, {van Bemmel}, {Br{\"u}ggen}, {Brunetti},
  {Conway}, {En{\ss}lin}, {Engels}, {Falcke}, {Ferrari}, {Haverkorn},
  {Jackson}, {Jarvis}, {Kapi{\'n}ska}, {Mahony}, {Miley}, {Morabito},
  {Morganti}, {Orr{\'u}}, {Retana-Montenegro}, {Sridhar}, {Toribio}, {White},
  {Wise}, \& {Zwart}}]{2016MNRAS.460.2385W}
{Williams}, W.~L., {van Weeren}, R.~J., {R{\"o}ttgering}, H.~J.~A., {et~al.}
  2016, \mnras, 460, 2385

\bibitem[{{Xue} \& {Wu}(2000)}]{2000ApJ...538...65X}
{Xue}, Y.-J. \& {Wu}, X.-P. 2000, \apj, 538, 65

\end{thebibliography}

\end{document}